\documentclass[twocolumn,secnumarabic,amssymb, nobibnotes, aps, prd]{revtex4}
\usepackage{graphicx}

\begin{document}
\title{The quantum mechanics is a non-universal theory. \\ The realistic Schrodinger's and positivistic Born's interpretation of the wave function}
\author{Alexey Nikulov}
\affiliation{Institute of Microelectronics Technology, Russian Academy of Sciences, 142432 Chernogolovka, Moscow District, Russia E-mail: nikulov@iptm.ru }
\begin{abstract} The controversies about quantum mechanics, in the old days and present-day, reveal an inconsistency of understanding of this most successful theory of physics. Therefore it is needed to set forth unambiguously what and how quantum mechanics describes in order to cut down the number of the fantasies trying to eliminate the fundamental obscurity in quantum mechanics. In this chapter reader's attention is drawn first of all to a non-universality of quantum-mechanical descriptions of different quantum phenomena. The realistic interpretation of the wave function proposed by Schrodinger is used at the description of most quantum phenomena whereas the controversies touch on the positivistic interpretation proposed by Born. These controversies are result, in the main, of the misinterpretation, proposed by Bohr, of the orthodox quantum mechanics. Most physicists, following Bohr, did not want to admit that Born had assumed in fact a mutual causal relation between quantum system and the mind of the observer. The EPR correlation is non-local and quantum mechanics predicts violation of the Bell's inequalities because of non-locality of the mind of the observer. The quantum postulate and complementarity proposed by Bohr are valid according to rather hidden-variables theories than the orthodox quantum mechanics. Measurement is described as process of interaction of quantum system with the measuring device in hidden-variables theories alternative of quantum mechanics. It is shown, that the mutual causal relation between 'res extensa' and 'res cogitans', presupposed with the Born's interpretation, results to a logical absurdity which testifies against the self-consistency of the orthodox quantum mechanic. This self-contradiction is a consequence of logical mistakes inherent in a new Weltanschauung proposed by Heisenberg for a philosophical substantiation of quantum mechanics. Quantum mechanics is successful in spite of this absurdity and these mistakes because rather the realistic Schrodinger's interpretation than the positivistic Born's interpretation is used at the description of the majority of quantum phenomena. The act of measurement and the fundamental obscurity connected with it are absent at this description. But there are other fundamental obscurities which are considered in the last section. 
\end{abstract}

\maketitle

\narrowtext

Contents \\

{\bf 1. Introduction}

{\bf 2. What is implied with the Born's interpretation of the wave function?} 

{\bf 2.1. 'Measurement' might be complete only in the mind of the observer}

{\bf 2.2. EPR correlation and the non-locality of the mind}

{\bf 2.3. Violation of Bell's inequalities uncovers the influence of subject on object}

{\bf 2.4. What EPR intended to prove and what they have proved}

{\bf 2.5. Hidden variables} 

{\bf 2.6. Indeterminism of quantum mechanics. The entanglement of cat with atom states} 

{\bf 2.7. Whose knowledge and whose will?}

{\bf 3. New Weltanschauung proposed by Heisenberg}

{\bf 3.1. Quantum mechanics rejects the Cartesian polarity between  'res cogitans' and  'res extensa'}

{\bf 3.2. The notion of the 'thing-in-itself' by Kant and hidden-variables} 

{\bf 3.3. The Kantian a priori character of the law of causality and quantum mechanics}

{\bf 4. Fundamental mistakes by sleepwalkers} 

{\bf 4.1. The quantum postulate and complementarity proposed by Bohr 'objectivate' observation}

{\bf 4.2. The mass delusion and the idea of quantum computation}

{\bf 4.3. Two principal mistakes of Heisenberg}

{\bf 5. Fundamental mistakes because of the prejudice of the QM universality} 

{\bf 5.1. The Aharonov - Bohm effects are described both with the $\psi $ - functions and the wave function} 

{\bf 5.2. We can believe for the time being in reality of the moon}

{\bf 6. Fundamental obscurity connected with wave function usage}

{\bf 6.1. Puzzles generated with the quantum formalism}

{\bf 6.2. Experimental results which can not be describe with help of the quantum formalism}

{\bf 7. Conclusion}

\section{Introduction}
Quantum mechanics (QM) is the most successful theory. It has given rise to revolutionary technologies of the XX century. The progress of physics of last century are fairly connected with the QM. But John Bell said in his Introductory remarks "Speakable and unspeakable in quantum mechanics" at Naples-Amalfi meeting, May 7, 1984 that "{\it This progress is made in spite of the fundamental obscurity in quantum mechanics}", see p. 170 in \cite{Bell2004}. This fundamental obscurity as well as QM are result of the proposal by young Werner Heisenberg \cite{Heisenberg1925} to describe observables instead of beables, see these terms in the Bell's paper \cite{Bell1976}. This proposal to abandon any attempt to find a unified picture of objective reality had provoked the battle between creators of quantum theory. Bohr, Pauli, Dirac and others admitted the Heisenberg's proposal whereas Einstein, Schrodinger, de Broglie and others rejected the repudiation of the science aim as the discovery of the real. Schrodinger interpreted his wave function as a real wave \cite{Schrodinger1926} and defended this realistic interpretation \cite{Schrodinger1952}. He tried to replace particles by wavepackets. {\it "But wavepackets diffuse"} \cite{Bell1987}. This diffuseness contradicts numerous observation. Therefore the interpretation of the Schrodinger's wave function as probability amplitudes proposed by Born was fully accepted by most physicists. This positivistic interpretation corresponds to the Heisenberg's proposal and just therefore it results to the fundamental obscurity and mass delusion. Indeterminism, subjectivity, non-locality and vagueness implied with this interpretation are enough obvious. But only few physicists, Einstein, Schrodinger, de Broglie and some others worried about these "philosophical" problems during a long time. Most physicists, as Bell said {\it "stride through that obscurity unimpeded... sleepwalking?"}, see p. 170 in \cite{Bell2004}. Bell worried about this obscurity of positivistic QM but even he said: {\it "The progress so made is immensely impressive. If it is made by sleepwalkers, is it wise to shout 'wake up'? I am not sure that it is. So I speak now in a very low voice"}, see p. 170 in\cite{Bell2004}. 

But now it is needed to shout "wake up" \cite{WakeUp}. There are some reasons why it is needed: 1) numerous false publications because of misunderstanding of QM \cite{WakeUp,Comment2010,Comment2009}; 2) misunderstanding of the idea of quantum computation \cite{Aristov2011}; 3) some authors, because of their implicit belief in QM, claim already that it is possible to prove experimentally that {\it "The moon - a small moon, admittedly - is not there"} \cite{Mooij2010}; 4) on the other hand many physicists have already refused this implicit belief  in QM. The most striking illustration of the fourth reason is the Action of the European Cooperation in Science and Technology "Fundamental Problems in Quantum Physics" \cite{COST}. The first aim of research - observer-free formulation of QM witnesses to perception by numerous participants of the Action MP1006 that no subjectivity can be permissible in any physical theory. But I should say that only this perception does not indicate that these scientists have 'waked up' completely. 

They have not realized for the present a primary logical mistake of the sleepwalkers creating QM. Scientists have in mind always that any physical theory should describe universally all its subject matters. For example, everyone believes that the Newton's laws describe universally the motion of all object with different mass, from major planets to smallest particles. This belief may be justified with a universality of the laws governing a unique objective reality. Orthodox QM, in contrast to all others theories of physics, describes rather different phenomena than a unique reality. No description of phenomena should be universal if they are not considered as a universal manifestation of a unique reality. Therefore it is logical mistake to think that QM should describe universally all quantum phenomena. Nevertheless QM was developed and interpreted up to now as a universal theory. General confidence predominates that only the positivistic Born's interpretation but not the realistic Schrodinger's interpretation can be valid for description of all quantum phenomena. 

This confidence is obviously false. Richard Feynman in the Section "The Schrodinger Equation in a Classical Context: A Seminar on Superconductivity" of his Lectures on Physics \cite{FeynmanL} stated that Schrodinger {\it "imagined incorrectly that $|\Psi |^{2}$ was the electric charge density of the electron. … It was Born who correctly (as far as we know) interpreted the $\Psi $ of the Schrodinger equation in terms of a probability amplitude…"}. But further Feynman wrote that {\it "in a situation in which $\Psi $ is the wave function for each of an enormous number of particles which are all in the same state, $|\Psi |^{2}$ can be interpreted as the density of particles"}. Thus, Feynman had pointed out that the positivistic Born's interpretation could be replaced with the realistic Schrodinger's interpretation at the description of macroscopic quantum phenomena, at least. This fact has fundamental importance because {\it "There are two fundamentally different ways in which the state function can change"} \cite{Everett1957} according to the Born's interpretation: the discontinuous change at the observation (Process 1 according to \cite{Everett1957}) and the continuous deterministic change of state of an isolated system with time according to a Schrodinger's wave equation (Process 2 according to \cite{Everett1957}). Hugh Everett noted correctly that because of the Process 1 {\it "No way is evidently be applied the conventional formulation of QM to a system that is not subject to external observation"} and that {\it "The question cannot be ruled out as lying in the domain of psychology"} \cite{Everett1957}. But only select few realized this fundamental obscurity in QM in that time. Feynman did not realized. Therefore he did not attach great importance to the replacement of the Born's interpretation by the Schrodinger's interpretation. Feynman did not understand that the Process 1, and all fundamental problems connected with it, disappear at this replacement. 

Richard Feynman and Hugh Everett were doctoral students of the same doctoral advisor - John Archibald Wheeler. But their conception of QM was fundamentally different. Such dissent marks out QM from other theories of physics. The dissent was from the very outset of QM. It was observed both between defenders of QM, for example Heisenberg and Bohr, and its critics, for example Schrodinger and de Broglie. But now the diversity of opinion is unusually wide. Einstein wrote as far back as 1928 to Schrodinger \cite{Einstein1928L}: {\it "The soothing philosophy-or religion?-of Heisenberg-Bohr is so cleverly concocted that it offers the believers a soft resting pillow from which they are not easily chased away"}, see the cite on the page 99 of \cite{QuCh2006}.  The diversity of opinion about QM witnesses that Einstein's words turned out prophetic: the dissent can be about a religion but our right comprehension must be unified. At least we must believe that it is possible. Otherwise no science could be possible. Therefore first of all it is important to show that subjectivity, non-locality, indeterminism and vagueness of QM are deduced unambiguously from the Born's interpretation. It will be made in the next Section. This positivistic interpretation can be valid and understood correctly only in a new Weltanschauung proposed by Heisenberg. This new Weltanschauung will be considered shortly in the Section 3. Unfortunately only few scientists have realized that the correct understanding of QM demands the new Weltanschauung. Both mass delusion connected with this lack of understanding and mistakes made by Heisenberg will be considered in the Section 4. Mistakes of other type connected with the misinterpretation of quantum mechanics an a universal theory will be considered in the Section 5. The fundamental obscurities in quantum mechanics worrying Einstein, Schrodinger, Bell and others disappear with the realistic interpretation of wave function proposed by Schrodinger. But other fundamental obscurities appear with this realistic interpretation. These fundamental obscurities of other type will be considered in the Section 6. 

\section{What is implied with the Born's interpretation of the wave function? }
Feynman wrote \cite{FeynmanL} when Schrodinger {\it "imagined incorrectly that $|\Psi |^{2}$ was the electric charge density of the electron… He soon found on doing a number of problems that it didn't work out quite right"}. Schrodinger had tried to replace 'particles' by wave-packets 
$$\Psi (r) = \int_{-\infty }^{\infty }dp[A_{0}(p)\exp{-\frac{i}{\hbar}Et}]\exp{\frac{i}{\hbar}pr} \eqno{(1)}$$                                                                          
But this wave-packets spreads in empty space, for example, when the energy $E = p^{2}/2m$. Therefore the real wave-packets cannot explain the observations of particle localized in the space, for example particle tracks in track chambers. Because of this and other defections of the Schrodinger's interpretation most physicists had accepted the Born's interpretation. Most of they stride unimpeded through the nonsense that the wave-packets can be localized only under influence of the mind of the observer.

\subsection{'Measurement' might be complete only in the mind of the observer}
Feynman wrote \cite{FeynmanL} that Born had proposed {\it "very difficult idea that the square of the amplitude is not the charge density but is only the probability per unit volume of finding an electron there, and that when you do find the electron some place the entire charge is there"}. Feynman was sure that this idea is correct because he did not raise the question: "How can the entire charge be there when an observer has found the electron some place?" Let consider the electron in empty space, or better a fullerene, or even a long biomolecule, quantum interference of which was observed already \cite{BiomolInt03,BiomolInt07}. Quantum state of such particle can be described with the wave-packets (1) in which the wave functions $exp i(pr - Et)/\hbar $ are deduced from the Schrodinger's wave equation \cite{LL}
$$i\hbar \frac{\partial  \Psi }{\partial t} = -\frac{\hbar ^{2}}{2m}\nabla ^{2}\Psi + U(r)\Psi  \eqno{(2)}$$  
whereas the amplitudes $A_{0}(p)$ can be estimate only with help of an observation at a time $t = 0$. According to (2) $E = p^{2}/2m$ in empty space where $U(r) = 0$. It is possible, for example with help of the scanning laser ionization detector used in \cite{Zeilinger02}, to observe at $t = 0$ that a fullerene, for example, is localised in a space region $\Delta r$ near $r = 0$. The result of this observation may be describe approximately with the probability density function $\Psi (r) = (2\pi )^{-1/4}\sigma  ^{-1/2}exp(-r^{2}/4\sigma ^{2})$ \cite{FeynmanL} where $\sigma \approx  \Delta r/3$. The wave-packet (1) diffuses, Fig.1, because its amplitudes, equal $A_{0}(p) = (8\pi )^{1/4}\sigma  ^{1/2}exp(-p^{2}\sigma ^{2}/\hbar ^{2})$ at $t = 0$, change with time $A(p,t) = (8\pi )^{1/4}\sigma  ^{1/2}exp(-p^{2}\sigma ^{2}/\hbar ^{2} - ip^{2}t/\hbar 2m)$. Because of this Process 2 \cite{Everett1957} the probability to observe the particle in a space region near $r = 0$ decreases and far off $r = 0$ increases with time, Fig.1. The Process 2 is continuous and is determined with the Schrodinger's wave equation (2). But the discontinuous change of the probability $|\Psi(r,t) |^{2}$ and the wave-packet $\Psi(r,t) $ at an observation, i.e. the Process 1 \cite{Everett1957}, can not be described with this equation (2). 

The Process 1, at $t = t_{1}$ for example, Fig.1, cannot be described outside the domain of psychology. Everett \cite{Everett1957} was right. No physical interaction of the particle with photons radiated with laser or any measuring device can compact the wave-packet. It is quite obvious that according to the Born's interpretation the wave-packet is squeezed in a smaller volume because of the observation by an observer the particle near, for example, $r \approx  8$, Fig.1. The observation changes first of all the mind of the observer. Before the observation at $t < t_{1}$ he conjectured to see the particle in any space region with a probability $|\Psi(r,t_{1}) |^{2}$. His knowledge changes discontinuously when he sees the particle near $r \approx  8$, Fig.1. Such change of the knowledge takes place at any observation. Schrodinger noted  that {\it"$\cdot \cdot \cdot$  the simple statement, that each observation depends both from the object and the subject which 'are entangled' by extremely complex manner is a statement which is hardly possible to consider new, it is old almost also, as the science"} \cite{SchrodingerHum}. But according to QM {\it "$\cdot \cdot \cdot$ the causal interconnection between the subject and object is considered reciprocal. It is stated, that the unremovable and uncontrollable influence of the subject on the object takes place"} \cite{SchrodingerHum}. According to the Born's interpretation both the observer knowledge and the quantum state change at the observation. 

\begin{figure}
\includegraphics{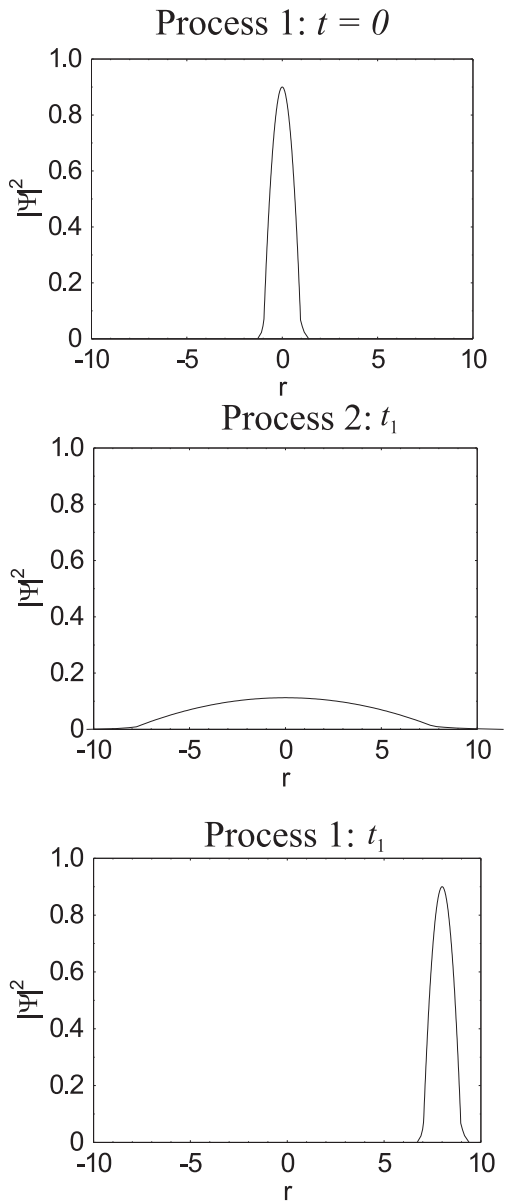}
\caption{\label{fig:epsart} The initial conditions, i.e. the amplitudes $A _{0}(p)$ of the wave-packet (1) are determined with results of an external observation at a time moment $t = 0$, i.e. during the Process 1, when the quantum state changes discontinuously under influence of an external observer (the upper picture). Quantum mechanics can predict the probability $|\Psi(r,t_{1}) |^{2}$ per unit volume of finding the particle in any space place $r$ at any time moment $t_{1}$ (the middle picture) using the initial conditions and the Schrodinger's wave equation (2). But no physical interaction of the particle with an agency of observation described with the Schrodinger's wave equation (2) can compact the wave-packet (the bottom picture) and provide with new initial conditions. The initial conditions can be provided only the mind of the observer who has found the particle some place.}
\end{figure}  

It could be clear from the very outset that according to the Born's interpretation the observation should be interpreted unambiguously as interplay between the quantum system and the mind of the observer. But Heisenberg and Bohr had convinced most physicists that we can consider the act of the observation (measurement) as an interaction between quantum system and measuring instrument. The famous \cite{QuCh2006} Heisenberg uncertainty microscope \cite{Heisenberg1927}, the quantum postulate and complementarity by Bohr \cite{Bohr1928} have misled some generations of physicists. Even the famous EPR paper \cite{EPR1935} and the Bell's works \cite{Bell2004} could not undeceive most physicists about this error up to now. Bell wrote in 1989 \cite{EPR1935} about the paper 'Ten theorems about quantum mechanical measurements', by NG van Kampen \cite{Kampen1988} {\it "This paper is distinguished especially by its robust common sense. The author has no patience with '$\cdot \cdot \cdot$ such mind-boggling fantasies as the many world interpretation $\cdot \cdot \cdot$'. He dismisses out of hand the notion of von Neumann, Pauli, Wigner - that 'measurement' might be complete only in the mind of the observer: '. . . I find it hard to understand that someone who arrives at such a conclusion does not seek the error in his argument'"}. There is important to remind that Everett had proposed the many world interpretation in order to describe the Process 1, i.e. 'measurement', as lying outside the domain of psychology \cite{Everett1957}. But the believers in the soothing philosophy or religion of Heisenberg-Bohr reject flatly, as well as van Kampen \cite{Kampen1988}, both the mind of the observer and the many world interpretation. Although it must be obvious that the attempt by Heisenberg and Bohr to propose the realistic substantiation of the uncertainty principle was false it is needed to explain again and again that EPR correlation and Bell's inequalities have proved this obvious fact. 
 
\subsection{EPR correlation and the non-locality of the mind}
Bell wrote in 1981 \cite{Bell1981}: {\it "The philosopher in the street, who has not suffered a course in quantum mechanics, is quite unimpressed by Einstein-Podolsky-Rosen correlations. He can point to many examples of similar correlations in everyday life. The case of Bertlmann's socks is often cited. Dr. Bertlmann likes to wear two socks of different colours. Which colour he will have on a given foot on a given day is quite unpredictable. But when you see that the first sock is pink you can be already sure that the second sock will not be pink"}.  The last sentence describes the influence of the object (the first sock) on the subject (the mind of the observer). Such influence can astonish nobody because {\it "it is old almost also, as the science"} \cite{SchrodingerHum}. Bell asked: {\it "And is not the EPR business just the same?"} \cite{Bell1981}. He considered a particular version of the EPR paradox \cite{EPR1935}, developed by David Bohm \cite{Bohm1951}, i.e. the Einstein-Podolsky-Rosen-Bohm gedanken experiment with two spin 1/2 particles. Quantum mechanics describes the spin states of two separate particles with two separate equations 
$$\psi _{A} = \alpha  _{A}|\uparrow  _{A}(r_{A})> +  \beta  _{A}|\downarrow  _{A}(r_{A})> $$ $$ \psi _{B} = \alpha  _{B}|\uparrow  _{B}(r_{B})> +  \beta  _{B}|\downarrow  _{B}(r_{B})> \eqno{(3)}$$
where the probability amplitudes $\alpha _{A}$, $\beta  _{A}$, $\alpha _{B}$, $\beta  _{B}$ depend on a free choice of a certain axis along which the component of particle spin will be measured and  $|\alpha _{A}|^{2} + |\beta  _{A}|^{2} = 1$, $|\alpha _{B}|^{2} + |\beta  _{B}|^{2} = 1$ always. The spin state of pair of two separate particle may be described also with the product of the equations (3) in which the amplitudes $\gamma _{1} = \alpha _{A} \alpha _{B}$, $\gamma _{2} =  \alpha _{A}\beta_{B}$, $\gamma _{3} = \beta_{A} \alpha _{B}$, $\gamma _{4} = \beta_{A}\beta_{B}$ and as a consequence $\gamma _{1}\gamma _{4} = \alpha _{A} \alpha _{B}\beta_{A}\beta_{B} = \gamma _{2}\gamma _{3} = \alpha _{A}\beta_{B} \beta_{A} \alpha _{B}$. The EPR correlation takes place when $\gamma _{1}\gamma _{4} \neq  \gamma _{2}\gamma _{3} $ and the spin states of the particles cannot be separated. Two particles in the singlet spin state 
$$\psi _{EPR} = \gamma _{2}|\uparrow  _{A}(r_{A})\downarrow _{B}(r_{B})> +  \gamma _{3}|\downarrow  _{A}(r_{A}) \uparrow_{B}(r_{B})> \eqno{(4)}$$
when $\gamma _{1} = 0$, $\gamma _{4} = 0$, and $\gamma _{2} \neq 0$, $\gamma _{3} \neq 0$ is called EPR pair. The axis along which the component of particle spin will be measured is the direction of a non-uniform magnetic field produced by magnets of a Stern-Gerlach analyzer \cite{QuCh2006}.  The particles will deflect up in the state $|\uparrow >$ and down in the state $|\downarrow>$. The observers $A$ (Alice) and $B$ (Bob) can choose any direction of their analyzer's axis. Whether either particle separately goes up or down on a given occasion is quite unpredictable. But according to the basic principle of quantum mechanics formulated by Dirac as far back as 1930 \cite{Dirac1930} {\it "$\cdot \cdot \cdot $ a measurement always causes the system to jump into an eigenstate of the dynamical variable that is being measured"}.  This Dirac jump, wave function collapse \cite{Neumann1932}, or {\it "'quantum jump'… from the 'possible' to the 'actual'"} \cite{Heisenberg1959} must take place logically during the act of observation because Alice cannot see that one particle deflects up and down simultaneously. When Alice sees that the particle deflects up in her Stern-Gerlach analyzer directed along an axis $n$ the spin state of the EPR pair (4) changes discontinuously to the eigenstate
 $$\psi _{n} = |\uparrow  _{A}(r_{A})\downarrow _{B}(r_{B})>_{n}   \eqno{(5a)}$$
or to the eigenstate 
$$\psi _{n} = |\downarrow  _{A}(r_{A}) \uparrow_{B}(r_{B})>_{n} \eqno{(5b)}$$
when the particle deflects down. Thus, according to the basic principles of quantum mechanics deduced logically from the Born's interpretation one particle $A$ of the EPR pair (4) goes up the other $B$ always goes down and vice-versa when axis of two analyzers located at widely separated points in space $r_{A}$ and $r_{B}$ is directed in the same direction $n$. 
This EPR correlation should be non-local because of the opportunity to observe the deflection of the particle $A$ and the particle $B$ at the same time independently on the distance $|r_{A} - r_{B}|$ between the Stern-Gerlach analyzers. This non-locality is deduced unambiguously from the Born's interpretation as a consequence of non-locality of the mind. Heisenberg noted: {\it "Since through the observation our knowledge of the system has changed discontinuously, its mathematical representation also has undergone the discontinuous change and we speak of a 'quantum jump'"} \cite{Heisenberg1959}. He justified the 'quantum jump' with help of the fact that {\it "our knowledge can change suddenly"} \cite{Heisenberg1959}, i.e. the obvious fact that the knowledge of Alice, for example, changes at the influence of the object, the deflection of the particle, on the subject, her mind. Already before the observation the Alice knowledge about the probability of the deflection up or down of particles $A$ and $B$ is entangled. She knows that if the deflection of one particle, for example $A$, will be up than the deflection of other particle $B$ should be down. Thus, the EPR correlation (4) describes the entanglement of the Alice knowledge about the spin state of two particles. Therefore, motivated \cite{entanglement2001} by EPR \cite{EPR1935}, Schrodinger coined \cite{Schrod35D,Schrod35D} the term {\it "entanglement of our knowledge"}: {\it "Maximal knowledge of a total system does not necessarily include total knowledge of all its parts, not even when these are fully separated from each other and at the moment are not influencing each other at all"}. Maximal knowledge of a total system include total knowledge of all its parts in the case (3) when $\gamma _{1}\gamma _{4} =  \gamma _{2}\gamma _{3} $ but not in the case (4) when $\gamma _{1}\gamma _{4} \neq  \gamma _{2}\gamma _{3} $.

In his last talk \cite{Bell1990} Bell considered the question: {\it "What can not go faster than light?"} He said: {\it "The situation is further complicated by the fact that there are things which do go faster than light. British sovereignty is the classical example. When the Queen dies in London (may it long be delayed) the Prince of Wales, lecturing on modern architecture in Australia, becomes instantaneously King, (Greenwich Mean Time rules here)"} \cite{Bell1990}. There is important to define more exactly that the Prince of Wales becomes instantaneously King in the mind of witnesses of the Queen death. Like manner the spin state of the distant particle $B$ changes instantaneously from (4) to (5a) in the mind of Alice when she sees that her particle $A$ has deflected up. The analogue of the British sovereignty in the EPR correlation is the Dirac jump, or wave function collapse, which should be at the Process 1. But there is a fundamental difference of the Dirac jump from the British sovereignty. The witnesses of the Queen death cannot have an influence on the Prince of Wales whereas Alice can govern the spin state of the distant particle $B$. Each of the eigenstates (5a) and (5b) of the operator of the spin component along $n$ (this dynamical variable) is superposition of eigenstates of the spin component along other direction (other dynamical variable) \cite{LL}. Suppose that initially Alice had chosen to orient the non-uniform magnetic field of her Stern-Gerlach analyzer perpendicular to the line of flight of the approaching particle, the y-axis, and pointing vertically upward along the z-axis. And Bob had oriented his non-uniform magnetic field also perpendicular to the y-axis but at an angle $\varphi $ to the z-axis. Than the EPR pair jumps discontinuously to the state
$$\psi _{z} = |\uparrow  _{A,z}\downarrow _{B,z}> =  cos(\varphi /2)|\uparrow  _{A,\varphi }\downarrow _{B,\varphi }> +$$ $$ +sin(\varphi /2) |\downarrow  _{A,\varphi } \uparrow_{B,\varphi }> \eqno{(6a)}$$
when Alice will see that her particle has deflected up and to the state 
$$\psi _{z} = | \downarrow_{A,z} \uparrow _{B,z}> =  -sin(\varphi /2)|\uparrow  _{A,\varphi }\downarrow _{B,\varphi }> +$$ $$+ cos(\varphi /2)|\downarrow  _{A,\varphi } \uparrow_{B,\varphi }> \eqno{(6b)}$$    
when her particle has deflected down. The operator of the turning round the y-axis \cite{LL} is used hear and below. Each of these states is the eigenstate of the operator corresponding to the orientation on the Alice analyzer but it is superposition the eigenstates of the operator corresponding to the orientation on the Bob analyzer. Alice can turn her Stern-Gerlach analyzer on an angle $\theta $ during the flight of the particles of the EPR pair (4). This turn changes the spin states of both her and Bob's particle after her observation to 
$$\psi _{\theta } = |\uparrow  _{A,\theta }\downarrow _{B,\theta }> =  cos((\varphi -\theta )/2)|\uparrow  _{A,\varphi }\downarrow _{B,\varphi }>+$$ $$ +sin((\varphi - \theta )/2) |\downarrow  _{A,\varphi } \uparrow_{B,\varphi }> \eqno{(7a)}$$
if her particle has deflected up and to 
$$\psi _{\theta } = | \downarrow_{A,\theta } \uparrow _{B,\theta }> =  -sin((\varphi - \theta )/2)|\uparrow  _{A,\varphi }\downarrow _{B,\varphi }>+$$ $$ + cos((\varphi - \theta )/2)|\downarrow  _{A,\varphi } \uparrow_{B,\varphi }> \eqno{(7b)}$$
if it has deflected down. Thus, the EPR correlation reveals that quantum mechanics concedes that Alice can change instantaneously the quantum state of the distant particle with her will and her observation. According to the Born's interpretation she can act faster than light thanks to non-locality of her mind.

\begin{figure}
\includegraphics{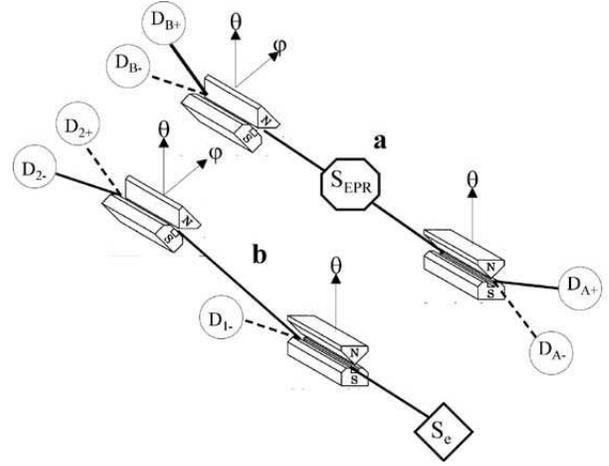}
\caption{\label{fig:epsart} Sketch of the EPR experiment (a) and experimental apparatus for measurement of  the probability $P_{\theta +}P_{\varphi -}$ with help of a source of single electrons â $S _{e}$ (b). In a case (a) one of the electrons of each EPR pair flies from the EPR pairs source $S_{EPR}$ to the Stern-Gerlach analyzer $A$, and another to the analyzer $B$. The probabilities $P_{A\theta +} = N_{A+}/(N_{A+} + N_{A-})$ and $P_{B\varphi  +} = N_{B+}/(N_{B+} + N_{B-})$ are defined as the relation of number $N_{A+}$ (or $N_{B+}$) of detection by detector $D_{A+}$ (or $D_{B+}$) to the sum $N_{A+} + N_{A-}$ (or $N_{B+} + N_{B-}$) of detection by detectors $D_{A+}$ and $D_{A-}$ (or $D_{B+}$ and $D_{B-}$). In a case (b) electron flies from the source $S _{e}$ to the first Stern-Gerlach analyzer and gets in the second analyzer if it deflects up, and gets in the first detector $D_{1-}$ if it deflects down. After the second analyzer electron gets in the detector $D_{2+}$ or $D_{2-}$. The probability $P_{\theta +}P_{\varphi -} = N_{2-}/(N_{1-} + N_{2+} + N_{2-})$ is defined as the relation of the number $ N_{2-}$ of detection by detector $D_{2+}$ to the sum $N_{1-} + N_{2+} + N_{2-}$ of detection by all detectors}
\end{figure} 

\subsection{Violation of Bell's inequalities uncovers the influence of subject on object}
Heisenberg justified \cite{Heisenberg1959} the discontinuous change conceded quantum mechanics with the argument that our knowledge of the system changes discontinuously at any observation. But quantum mechanics represents not only our knowledge. It predicts first of all the probability of different outcomes of observations. For example, the relation (7) predicts that the probability to observe the deflection up of the Bob's particle equals  $|cos((\varphi  - \theta )/2)|^{2}$ and down $|sin((\varphi  - \theta )/2)|^{2}$ when the Alice particle has deflected up. Thus, the Alice's will and her observation influence instantaneously on the outcome of observations of the distant particle. This influence is revealed most definitely with help of the Bell's inequalities \cite{Bell1964}. Only condition used at the deduction of the Bell's inequality is {\it "the requirement of locality, or more precisely that the result of a measurement on one system  be unaffected by operations on a distant system"} \cite{Bell1964}. Bell had proposed in \cite{Bell1981} a most simple example of this logical deduction. Bell started with an trivial inequality 
$$P_{0+}P_{45-} + P_{45+}P_{90-} \geq  P_{0+}P_{90-} \eqno{(8a)}$$
asserting that the probability $P_{0+}$ of the deflection up at the orientation of the Stern-Gerlach analyzer vertically upward along the z-axis, i.e. at $\theta  = 0^{o}$  and the $P_{45-}$  of the deflection down at $\theta  = 45^{o}$ plus the probability $P_{45+}$  of the deflection up at $\theta  = 45^{o}$ and $P_{90-}$  - down at $\theta  = 90^{o}$ is not less than the probability $P_{0+}$  of the deflection up at $\theta  = 0^{o}$ and $P_{90-}$  - down at $\theta  = 90^{o}$. The inequality is obvious when all probabilities $P_{0+}$, $P_{45-}$, $P_{45+}$  and $P_{90-}$ are measured in the same spin state. Any particle in the same spin state which deflects up at $\theta  = 0^{o}$ and down at $\theta  = 90^{o}$  (and so contributing to the third probability $P_{0+}P_{90-}$ in (8a)) can deflect either up at $\theta  = 45^{o}$ (and so contributes to the second probability $P_{45+}P_{90-}$ in (8a)) or down at $\theta  = 45^{o}$ (and so contributes to the first probability $P_{0+}P_{45-}$ in (8a)). The inequality is trivial but it can not be verified experimentally with measurements of single particles, as shown on Fig2b, because {\it "$\cdot \cdot \cdot $ a measurement always causes the system to jump into an eigenstate of the dynamical variable that is being measured"} \cite{Dirac1930}. All particles flying from the first to the second Stern-Gerlach analyzer on Fig.2b should be in the state 'spin up' because of this Dirac jump. The eigenstate 'spin up' for this orientation of the first Stern-Gerlach analyzer differs in common case from the initial spin state of the particles. Therefore it is impossible to measure the probabilities both $P_{0+}$ and $P_{45-}$, for example, in the same spin state with the method shown on Fig.2b. The probabilities at different orientation of the Stern-Gerlach analyzers $\theta $ and $\varphi $ can be measured  in the same state with help of the EPR pair if the requirement of locality is valid. The equality of the probabilities $P_{A\theta +} = P_{B\theta -}$ and $P_{A\theta -} = P_{B\theta +}$ must be observed because of the EPR correlation when if one particle $A$ of the EPR pairs (4) deflects up then the other $B$ always deflects down and vice-versa. This equality is valid for all orientation including $\theta  = 45^{o}$ and $\theta  = 90^{o}$. The Bell's inequality 
$$P_{A0+}P_{B45+} + P_{A45+}P_{B90+} \geq  P_{A0+}P_{B90+} \eqno{(8b)}$$
is deduced from the obvious inequality (8a) at the single requirement: a turning of the Stern-Gerlach analyzer $A$ located in a space region $r_{A}$ can not change instantaneously on the spin of the distant particle $B$ located in a space region $r_{B}$ and vice-versa. The probability to observe the deflection up in the Stern-Gerlach analyzers $A$ equals always the same value $P_{A\theta +} = (|\gamma _{2}|^{2} + |\gamma _{3}|^{2})/2 = 0.5$ if Alice is first observer because of the same probability for each particle of the EPR pair (4) to flow toward Alice. Bob, as second observer, should observe the deflection up with the probability $P_{B\varphi +} =|sin((\varphi  - \theta )/2)|^{2}$ according to (7a). The result $P_{A\theta +}P_{B\varphi +} =0.5|sin((\varphi  - \theta )/2)|^{2}$ gives the values $P_{A0+}P_{B45+} = P_{A45+}P_{B90+} =  0.5sin^{2}(45^{o}/2) \approx  0.0732$, $P_{A0+}P_{B90+} =  0.5sin^{2}(90^{o}/2) \approx  0.25$. The inequality (8b) would then require 
$$0.1464 \geq  0.25  \eqno{(8c)}$$
which is not true. 

This violation (8c) of the Bell's inequality (8b) reveals that quantum mechanics presupposes that a turning of the Stern-Gerlach analyzer can influence instantaneously on the distant particle because the Bell's inequality (8b) was deduced only from the requirement of impossibility of such non-local influence. According to the Born's interpretation this non-local influence is actualized by means of the Alice's mind. The probability that the Bob's particle will deflect up should equal $P_{B\varphi +} = 0.5$ until Alice has seen that her particle has deflected up. Thus the probability of the observation 'spin up' by Bob changes from $P_{B\varphi +} = 0.5$ to $P_{B\varphi +} =|sin((\varphi  - \theta )/2)|^{2}$ because of the discontinuous change of the Alice's knowledge. The knowledge changes because of the influence of object on subject and the probability of the observation changes because of the influence of subject on object. Thus, the EPR correlation and the Bell's inequalities have confirmed the statement by Schrodinger that in the orthodox quantum mechanics {\it "$\cdot \cdot \cdot $ the causal interconnection between the subject and object is considered reciprocal. It is stated, that the unremovable and uncontrollable influence of the subject on the object takes place"} \cite{SchrodingerHum}. 

\subsection{What EPR intended to prove and what they have proved}
EPR \cite{EPR1935} denied any possibility of the EPR correlation. It was in conflict with the Einstein's belief: {\it "But on one supposition we should, in my opinion, absolutely hold fast: the real factual situation of the system $S_{2}$ is independent of what is done with the system $S_{1}$, which is spatially separated from the former"} \cite{EinsteiAutob}. EPR intended to prove {\it "that the description of reality as given by a wave function is not complete"} \cite{EPR1935}. Of course they had in mind the wave function in the Born's interpretation. It is stated in the abstract of the EPR paper \cite{EPR1935}: {\it "In quantum mechanics in the case of two physical quantities described by non-commuting operators, the knowledge of one precludes the knowledge of the other. Then either (1) the description of reality given by the wave function in QM is not complete or (2) these two quantities cannot have simultaneous reality. Consideration of the problem of making predictions concerning a system on the basis of measurements made on another system that had previously interacted with it leads to the result that if (1) is false then (2) is also false"}. Indeed, if Alice reveals a real situation existing irrespective of any act of observation when she sees that the particle has deflected up in her Stern-Gerlach analyzer pointing vertically upward along the z-axis (6a) than all other spin components (7a) or (7b) of her and Bob's particles should exist before her observation. Alice can know any spin component (7a) or (7b) in the same spine state turning her analyzer in the respective axis. Therefore she can obtain the knowledge about different spin component of the Bob's particle, $|\downarrow _{B,z}>$ and $|\downarrow _{B,\theta }>$ for example, contrary to the foundation of QM, if this knowledge about a real situation existing irrespective of her mind. 

Thus, QM is observably inadequate if it is interpreted as the description of reality. The description given by the wave function in QM can be adequate and complete only if two physical quantities described by non-commuting operators does not have reality simultaneous before their observation. QM can be valid only if physical quantities are rather created by the mind of the observer than measured at the observation. Just this absurdity of QM had been proved by EPR \cite{EPR1935}. Bell proposed {\it "to replace the word 'measurement'"} which misleads: {\it "When it is said that something is 'measured' it is difficult not to think of the result as referring to some pre-existing property of the object in question"} \cite{EPR1935}. The pre-existing properties revealed at measurement are in hidden-variables theory, alternative the orthodox QM using the Born's interpretation. According to this theory, hidden variables determine results of individual measurements and thus eliminate subjectivity and indeterminism inherent QM.  

\subsection{Hidden variable}
Bell asserted {\it "that vagueness, subjectivity, and indeterminism, are not forced on us by experimental facts, but by deliberate theoretical choice"} \cite{Bell1982}. It is not absolutely so. The main reason of refusal of realism were problems with the realistic description of some quantum phenomena, such as Stern-Gerlach effect \cite{StGe1922}. Bohr wrote in 1949 \cite{Bohr1949}, that {\it "as exposed so clearly by Einstein and Ehrenfest \cite{Einstein1922}, it presented with unsurmountable difficulties any attempt at forming a picture of the behaviour of atoms in a magnetic field"}. And Bell wrote 32 years later: {\it "Phenomena of this kind made physicists despair of finding any consistent space-time picture of what goes on the atomic and subatomic scale $\cdot \cdot \cdot $ Going further still, some asserted that atomic and subatomic particles do not have any definite properties in advance of observation. There is nothing, that is to say, in the particles approaching the magnet, to distinguish those subsequently deflected up from those subsequently deflected down. Indeed even the particles are not really there"} \cite{Bell1981}. 

 As Bell wrote \cite{Bell1966} {\it "To know the quantum mechanical state of a system implies, in general, only statistical restrictions on the results of measurements"}. The classical statistical mechanics also describes only statistical distribution of parameters. But these parameters are assumed to exist irrespective of any act of observation and the mind of the observer. Just the negation of real existence of parameters results to subjectivity and indeterminism of QM. Therefore Bell was sure: {\it "It seems interesting to ask if this statistical element be thought of as arising, as in classical statistical mechanics, because the states in question are averages over better defined states for which individually the results would be quite determined. These hypothetical 'dispersion free' states would be specified not only by the quantum mechanical state vector but also by additional 'hidden variables'-'hidden' because if states with prescribed values of these variables could actually be prepared, quantum mechanics would be observably inadequate"} \cite{Bell1966}. But most physicists did not take an interest in this problem fifty years ago and up to now many physicists underestimate the fundamental importance of hidden variables. Few experts, who did find the question interesting, believed that {\it "the question concerning the existence of such hidden variables received an early and rather decisive answer in the form of von Neumann's proof on the mathematical impossibility of such variables in quantum theory"} \cite{Bell1966}. 

But this belief was false. David Mermin writes in the paper "Hidden variables and the two theorems of John Bell" \cite{Mermin1993}: {\it "A third of a century passed before John Bell, 1966, rediscovered the fact that von Neumann's no-variables-hidden proof was based on an assumption that can only be described as silly - so silly, in fact, that one is led to wonder whether the proof was ever studied by either the students or those who appealed to it"}. Von Neumann did not take into account that non-commuting operators do not have simultaneous eigenvalues \cite{Mermin1993}. There is more important to realize a physical mistake corresponding to this 'mathematical' mistake. Eigenvalues non-commuting operators cannot be simultaneous measured according to the quantum postulate and complementarity proposed by Bohr \cite{Bohr1928}. Bell noted \cite{Bell1966} that additional demands of the von Neumann's proof are {\it "quite unreasonable when one remembers with Bohr \cite{Bohr1949} 'the impossibility of any sharp distinction between the behaviour of atomic objects and the interaction with the measuring instruments which serve to define the conditions under which the phenomena appear'"}. In order to exhibit of the error of the von Neumann's proof Bell had constructed a hidden-variables model which reproduces all predictions of results of a single spin 1/2 measurement given by QM. There is important to accentuate that Bell used in this model the Bohr's quantum postulate which {\it "implies that any observation of atomic phenomena will involve an interaction with the agency of observation not to be neglected"} \cite{Bohr1928}. According to the quantum postulate {\it "With or without hidden variables the analysis of the measurement process presents peculiar difficulties"} \cite{Bell1966}, because no theory can describe an interaction with the agency of observation. Any result of this interaction may be assumed because of this vagueness. 

In the Bell's model \cite{Bell1966} the interaction of single spin 1/2 with the agency of observation results to measurement of the same value of spin component, as it is observed in the paradoxical  Stern-Gerlach effect \cite{StGe1922}. Thanks to the vague interaction the results of observation can be described with the relation $s_{n} = 1/2 cos(\theta )/|cos(\theta )|$ proposed by Bell in \cite{Bell1981}. This relation describes an individual measurement of spin component $s_{n}$ along $n$, as well as the superposition (3), but has a fundamental advantage: $\theta $ is the angle between an axis ${\bf n}$ of Stern-Gerlach analyzer and an axis ${\bf z} + {\bf m}$ of spin. Therefore results of an individual measurement is determined with the spin axis ${\bf z} + {\bf m}$ and the mind of the observer can not influence on these results. In the Bell's model \cite{Bell1966} ${\bf z}$ is a unit vector directed along the z-axis in the spin state $\psi _{z} = |\uparrow _{z}>$ and ${\bf m}$ is a random unit vector which plays the role of hidden variable. This model predicts the same probabilities of observation of positive values $P_{\theta +} = 1/2 + \int _{\theta }^{\pi /2}dx2\pi \sin x/4\pi = (1 + \cos \theta )/2 = cos ^{2}(\theta /2)$ and negative values $P_{\theta -} = \int _{0}^{\theta }dx2\pi \sin x/4\pi = (1 - \cos \theta )/2 = sin ^{2}(\theta /2)$ as orthodox QM. 

It predicts also that if one particle $A$ of the EPR pair goes up the other $B$ always goes down and vice-versa when axis of two analyzers is directed in the same direction $n$ because if $cos(\theta _{A}) > 0$ for the angle $\theta _{A}$ between ${\bf n}$ and $+({\bf z} + {\bf m})$ then $cos(\theta _{B}) < 0$ for the angle $\theta _{B}$ between ${\bf n}$ and $-({\bf z} + {\bf m})$ and vice-versa. The hidden-variables model implies also the Dirac jump at observation in order to correspond to the prediction of QM. Therefore it also predicts violation of the trivial inequality (8a). Because of the vague interaction with the agency of observation, i.e. with the first Stern-Gerlach analyzer, shown on Fig.2b, the particle deflecting with the probability $P_{\theta +} = 0.5$ to the second Stern-Gerlach analyzer jumps to the spin state $\psi _{\theta } = |\uparrow _{\theta }>$. Therefore the particle will deflect down after the second analyzer with the probability $P_{\varphi -} =|sin((\varphi  - \theta )/2)|^{2}$. The total probability $P_{\theta +}P_{\varphi -} =0.5|sin((\varphi  - \theta )/2)|^{2}$ to hit the detector $D_{2-}$ predicts violation (8c) of the trivial inequality (8a): $P_{0+}P_{45-} = P_{45+}P_{90-} =  0.5sin^{2}(45^{o}/2) \approx  0.0732$, $P_{0+}P_{90-} =  0.5sin^{2}(90^{o}/2) \approx  0.25$. Thus, the hidden-variables model can reproduce almost all prediction of QM. Among few predictions which it can not reproduce is violation of the Bell's inequality (8b). The probability of observation both $s_{n} = +1/2$  and $s_{n} = -1/2$ of each particle of the EPR pairs is the same $P_{A\theta +} = P_{A\theta -} = P_{B\varphi +} = P_{B\varphi -} = 0.5$ because the results of individual measurements are determined by the spin axis $+({\bf z} + {\bf m})$ or $-({\bf z} + {\bf m})$ and therefore the act of measurement of one particle can not influence on the result of observation of other particle. The corroboration of the Bell's inequality (8b) $P_{A0+}P_{B45+} + P_{A45+}P_{B90+} =  0.5 \times 0.5  + 0.5 \times 0.5 = 0.5 > P_{A0+}P_{B90+} =  0.5 \times 0.5 =  0.25$ reveals that if definite properties exist in advance of observation then 'measurement' might be complete without the mind of the observer. 

There is important to note that hidden variables replace the mind of the observer with soulless agencies of observation. This fact reveals that the quantum postulate and complementarity proposed by Bohr \cite{Bohr1928} is valid according to rather hidden-variables theories than the QM based on the Born's interpretation. Variables are hidden just because {\it "any observation of atomic phenomena will involve an interaction with the agency of observation not to be neglected"} \cite{Bohr1928}. Bohr concluded from this statement implied with his quantum postulate that {\it "an independent reality in the ordinary physical sense can neither be ascribed to the phenomena nor to the agencies of observation"} \cite{Bohr1928}. This conclusion misleads. A celebrated polymath who is quoted in \cite{Mermin1993} declared that {\it "Most theoretical physicists are guilty of $\cdot \cdot \cdot $ fail[ing] to distinguish between a measurable indeterminacy and the epistemic indeterminability of what is in reality determinate. The indeterminacy discovered by physical measurements of subatomic phenomena simply tells us that we cannot know the definite position and velocity of an electron at any instant of time. It does not tell as that the electron, at any instant of time, does not have a definite position and velocity. [Physicists] $\cdot \cdot \cdot $ convert what is not measurable by them into the unreal and the non-existent"} \cite{Adler1992}. Bohr was among these most theoretical physicists and had misled some generation of physicists with his quantum postulate and complementarity. He did not take into account that an interaction with the agency of observation can change variables at observation but it can not create observed variables. Only the mind of the observer can create definite properties observed by the observer if they were not definite with variables even hidden in advance of observation. The EPR correlation and violation (8c) of the Bell's inequality (8b) reveal that an interaction rather with the mind of the observer than with the agency of observation is implied in the orthodox QM thanks to the non-locality of the first interaction and the locality of the second one.

\subsection{Indeterminism of quantum mechanics. The entanglement of cat with atom states }
Von Neumann, Pauli, Wigner and also Heisenberg and others were forced to note {\it "that 'measurement' might be complete only in the mind of the observer"} \cite{EPR1935} because of indeterminism of QM: if a cause of a definite result of the observation is absent in nature then only the mind of the observer can be the cause. Bohr reminder in 1949 \cite{Bohr1949} that during the Solvay meeting 1927 {\it "interesting discussion arose also about how to speak of the appearance of phenomena for which only predictions of statistical character can be made. The question was whether, as to the occurrence of individual effects, we should adopt a terminology proposed by Dirac, that we were concerned with a choice on the part of "nature" or, as suggested by Heisenberg, we should say that we have to do with a choice on the part of the 'observer' constructing the measuring instruments and reading their recording"}. Bohr wrote: {\it "Any such terminology would, however, appear dubious since, on the one hand, it is hardly reasonable to endow nature with volition in the ordinary sense, while, on the other hand, it is certainly not possible for the observer to influence the events which may appear under the conditions he has arranged"} and advertised his complementarity: {\it "To my mind, there is no other alternative than to admit that, in this field of experience, we are dealing with individual phenomena and that our possibilities of handling the measuring instruments allow us only to make a choice between the different complementary types of phenomena we want to study"} \cite{Bohr1949}. But this pet idea by Bohr can not answer on the question: "What or who can a definite result of the observation determine?" It can result and even had resulted \cite{LL} to the illusion that individual effects are chosen by the agency of observation, named  {\it "the 'classical object' usually called apparatus"} \cite{LL}. 

This illusion is logically absurd. Nevertheless it predominated and predominates up to now among most physicists thanks to the followers \cite{LL} of Bohr. Bell had called the spontaneous jump of a 'classical' apparatus into an eigenstate of its 'reading' as the LL jump, considering in \cite{EPR1935} Quantum Mechanics by L D Landau and E M Lifshitz \cite{LL} as the first of the 'good books' which mislead. The assumption \cite{LL} about the LL jump is absurd first of all because a apparatus even 'classical' is a part of nature as well as the cat in the famous paradox proposed by Schrodinger \cite{Schrod35D}. The Schrodinger's cat paradox is well-known but pure understood. Therefore it is useful to reminder its text here: {\it "One can even set up quite ridiculous cases. A cat is penned up in a steel chamber, along with the following diabolical device (which must be secured against direct interference by the cat): in a Geiger counter there is a tiny bit of radioactive substance, so small that perhaps in the course of one hour one of the atoms decays, but also, with equal probability, perhaps none; if it hap-pens, the [Geiger] counter tube discharges and through a relay releases a hammer which shatters a small flask of hydrocyanic acid. If one has left this entire system to itself for an hour, one would say that the cat still lives if meanwhile no atom has decayed. The first atomic decay would have poisoned it. The $\psi $-function of the entire system would express this by having in it the living and the dead cat (pardon the expression) mixed or smeared out in equal parts"}  \cite{Schrod35D}, see also p.185 in the book \cite{QuCh2006}. 
	Schrodinger had entangled with the $\psi $-function of the entire system 
$$\Psi_{cat} = \alpha At_{decay}G_{yes}Fl_{yes}Cat_{dead}  +$$ $$ + \beta At_{no} G_{no}Fl_{no}Cat_{living}   \eqno{(9)} $$ 
cat state $Cat_{dead}$, $Cat_{living}$ with the states of the small flask of hydrocyanic acid $Fl_{yes}$, $Fl_{no}$, the Geiger counter tube $G_{yes}$, $G_{yes}$ and radioactive atom $At_{decay}$, $At_{no}$ with the experiment conditions. The act of observation of the dead cat is described with the $\psi $ - function (9) collapse to 
$$\Psi_{cat} = At_{decay}G_{yes}Fl_{yes}Cat_{dead} \eqno{(10)} $$
One can draw the conclusion that the cat is dead $Cat_{dead}$ because the hammer has shattered the small flask of hydrocyanic acid $Fl_{yes}$. The hammer has shattered it because the Geiger counter tube has discharged $G_{yes}$. It is has discharged because the atom has decayed $At_{decay}$. Till this each event had a cause. But the atom decay is causless. There is no term to the right of $At_{decay}$ in (9). According to the assumption \cite{LL} about the LL jump the cat kills himself. Moreover one may demonstrate to combine two famous paradoxes, the EPR paradox and the Schrodinger's cat paradox, that the death of a cat $A$ can preserve life of a distant cat $B$ and vice-versa. Thereto one may substitute of the radioactive atom for the EPR pairs with two spin particles in the singlet state (4), as well as in the Bohm's version \cite{Bohm1951} of the EPR paradox. We will use also two Stern-Gerlach analysers, two Geiger counter tubes, two flasks of hydrocyanic acid and two cats $Cat_{A}$ and $Cat_{B}$. The Geiger counter tubes will be located on the upper trajectory of each particle after its exit from its Stern-Gerlach analyser, so it will discharge when spin up and will not discharge when spin down. The subsequent events will be as well as in the Schrodinger paradox \cite{Schrod35D}. This gedankenexperiment can be described with the $\psi $-function 
$$\Psi_{EPR,cat} = \gamma _{2} |\uparrow _{A}(r_{A})\downarrow _{B}(r_{B})> Cat_{A,dead}Cat_{B,liv}  + $$ $$ + \gamma _{3} |\downarrow _{A}(r_{A})\uparrow _{B}(r_{B})> Cat_{A,liv} Cat_{B,dead}   \eqno{(11)} $$
with two types of entanglements: because of the conservation law (the EPR correlation) and because of the condition of experiment proposed by Schrodinger \cite{Schrod35D}. The results of observations will be 
$$\Psi_{EPR,cat} = Cat_{A,dead}Cat_{B,liv}   \eqno{(12a)} $$          
or 
$$\Psi_{EPR,cat} = Cat_{A,liv} Cat_{B,dead}   \eqno{(12b)} $$
when the axis of the Stern-Gerlach analysers are parallel. Advocates of quantum  mechanics justify the using of state superposition (11) with the absence of any cause of the atom decay. They {\it "convert what is not measurable by them into the unreal and the non-existent"} \cite{Adler1992}.  Let imagine that we do not know why the observed states of cats are correlated as well as we do not know the cause of atom decay. Then, to convert our lack of knowledge into the unreal we can describe the results of our observations of the cats state with superposition    
$$\Psi_{EPR,cat} = \gamma _{2} Cat_{A,dead}Cat_{B,liv}  + \gamma _{3} Cat_{A,liv} Cat_{B,dead}   \eqno{(13)} $$
which should collapse to (12a) or (12b) at each observation. The absence of any cause of (12a) or (12b) in advance of observation raises a question: "What or who makes a choice?" According to the concept of the spontaneous collapse of a macroscopic system into a definite macroscopic configuration \cite{LL}, i.e. the LL jump \cite{EPR1935}, the choice is made by a cat, which is a 'classical' apparatus in the Schrodinger's paradox. But what cat $A$ or $B$ would make a choice (12a) or (12b)? The collapse of the superposition of cats states (13)  must be instantaneous irrespective of a distance $|r_{A} - r_{B}|$ between cats. According to the principle of relativity by Einstein both the cat $A$ and the cat $B$ may be observed first in the same case but in different frames of reference. Therefore it is impossible to say the spontaneous collapse of what cat can choose the fate of other cat with help a mystical action of a distance. The absurdity of the LL jump assumed in \cite{LL} is obvious even without this consideration of the cats fate. It must be obvious for any one that only a magical apparatus even 'classical' can collapse spontaneously.  

\subsection{Whose knowledge and whose will?}
Therefore von Neumann, Pauli, Wigner, Heisenberg and others admitted {\it "that 'measurement' might be complete only in the mind of the observer"} and the Dirac jump is forced by an external intervention \cite{EPR1935} which can be only the mind of the observer. But a choice on the part of the 'observer' suggested by Heisenberg can not deliver from a logical absurdity. According to the basic principle of QM formulated by Dirac {\it "$\cdot \cdot \cdot $ a measurement always causes the system to jump into an eigenstate of the dynamical variable that is being measured"} \cite{Dirac1930}. But into which eigenstate should the system jump when two dynamical variables described by non-commuting operators are measured at the same time? Alice may orient her Stern-Gerlach analyser at an angle $\theta $ to the z-axis and Bob may orient his Stern-Gerlach analyser at an other angle $\varphi $. Then Alice will be sure that her and Bob's particles jump to the spin state (7a) when she will see that her particle has deflected up. But Bob will be sure that his and Alice's particles jump to the other state
$$\psi _{\varphi } = | \downarrow_{A,\varphi } \uparrow _{B,\varphi }> =  -sin((\theta - \varphi )/2)|\uparrow  _{A,\theta }\downarrow _{B,\theta }> + $$ $$ + cos((\theta -\varphi )/2)|\downarrow  _{A,\theta } \uparrow_{B,\theta }> \eqno{(14)}$$
when he will see that his particle has deflected up. Thus, according to orthodox QM, knowledge of two observers of the same system can be various. Moreover, each of them can impose her (or his) will on the distant particle. When Alice has oriented her Stern-Gerlach along the z-axis the Bob's particle should jump in the spin state (5a) or (5b) at her measurement. And when Bob has oriented his Stern-Gerlach at an angle $\theta $ to the z-axis the Alice's particle should jump in the spin state (6a) or (6b) at his measurement. Here it is impossible to solve, whose knowledge is correct, and whose will can win because of the principle of relativity according to which Alice observes her particle ahead of Bob in a frame of reference whereas in an other frame of reference Bob observes his particle ahead of Alice. 

This absurdity of QM has became especially relevant after experimental evidence \cite{Aspect1981,Aspect1982,Aspect82} of violation of the Bell's inequalities. Before these experiments Bell expressed a hope that {\it "Perhaps Nature is not so queer as quantum mechanics"} \cite{Bell1981} and rated a possibility of violation of his inequalities as indigestible.  One of the interpretations of this violation could be a conclusion that {\it "Apparently separate parts of the world would be deeply and conspiratorially entangled, and our apparent free will would be entangled with them"} \cite{Bell1981}. The results of the experiments of the Aspect's team \cite{Aspect1981,Aspect1982,Aspect82} Bell appraised as a fundamental problem of theory: {\it "For me then this is the real problem with quantum theory: the apparently essential conflict between any sharp formulation and fundamental relativity. That is to say, we have an apparent incompatibility, at the deepest level, between the two fundamental pillars of contemporary theory"} p. 172 in \cite{Bell2004}. As opposed to Bell contemporary believers in the soothing philosophy or religion of Heisenberg-Bohr, for example the authors of the book \cite{Nielsen2000} are sure that violation of the Bell's inequalities has corroborated the correctness of QM. Mermin wrote as far back as in 1985 \cite{Mermin1985}: {\it "In the question of whether there is some fundamental problem with quantum mechanics signaled by tests of Bell's inequality, physicists can be divided into a majority who are 'indifferent' and a minority who are 'bothered'"}. 

This division observed up to now witnesses against QM as a consistent and transparent theory. The inconsistency the assessment discloses vagueness of QM. The majority who are 'indifferent' rather believe than understand QM. The authors of the book \cite{Nielsen2000} and other believers do not want to understand that no experimental result can save QM because it is self-contradictory and vague. Because of its vagueness the contradictions were observed even between its creators, Heisenberg and Bohr, first of all about the role of the observer. Heisenberg admitted that 'measurement' might be complete only in the mind of the observer. It is obvious, for example, from his destructive criticism of Soviet scientists Alexandrov and Blochinzev in the Section VIII "Criticism and Counterproposals to the Copenhagen Interpretation of Quantum Theory" of the Lectures 1955-1956 "Physics and Philosophy" \cite{Heisenberg1959}. These Soviet scientists stated that {\it "Among the different idealistic trends of contemporary physics the so-called Copenhagen school is the most reactionary"} \cite{Heisenberg1959} and rejected the role of the observer in QM. Heisenberg quotes Alexandrov {\it "We must therefore understand by 'result of measurement' in quantum theory only the objective effect of the interaction of the electron with a suitable  object. Mention of the observer must be avoided, and we must treat objective conditions and objective and effects. A physical quantity is objective characteristic of phenomenon, but not the result of an observation"} and notes {\it "According to Alexandrov, the wave function in configuration space characterizes the objective state of the electron"} \cite{Heisenberg1959}. Further Heisenberg explains why the Alexandrov's point of view is false: {\it "In his presentation Alexandrov overlooks the fact that the formalism of quantum theory does not allow the same degree of objectivation as that of classical physics. For instance, if a interaction of a system with the measuring apparatus is treated as a whole according to QM and if both are regarded as cut off from the rest of the world, then the formalism of quantum theory does not as a rule lead to a define result; it will not lead, e.g., to the blackening of the photographic plate in a given point. If one tries to rescue the Alexandrov's 'objective effect' by saying that 'in reality' the plate is blackened at a given point after the interaction, the rejoinder is the quantum mechanical treatment of the closed system consisting of electron, measuring apparatus and plate is no longer being applied"} \cite{Heisenberg1959}. 

This disproof by Heisenberg of the objectivation of QM is doubtless. Its obviousness is illustrated in the Section 2.1 and must be quite clear at consideration of the example shown on Fig.1: the wave-packet can be compacted only by the mind of the observer. In spite of this obviousness not only the Soviet scientists Alexandrov and Blochinzev but most physicists including Bohr objectified and objectify the matter of quantum mechanical treatment. Bohr objectified it with his quantum postulate and complementarity \cite{Bohr1928} according to which the act of observation is an interaction with the agency of observation. Most physicists had followed rather Bohr than Heisenberg because of their robust common sense according to which both the many world interpretation and the mind of the observer are mind-boggling fantasies. QM seems reasonable to most physicists only thanks to its misinterpretation. The following word by Heisenberg can explain partly the cause of this mass delusion: {\it "Above all, we see from these formulations how difficult it is when we try to push new ideas into an old system of concepts belonging to an earlier philosophy - or, to use an old metaphor, when we attempt to put new wine into old bottles"} \cite{Heisenberg1959}. Therefore it is needed to give an account of the essence of a new bottle proposed by Heisenberg for QM.

\section{New Weltanschauung proposed by Heisenberg}  
QM had originated from the proposal by young Heisenberg {\it "to try to establish a theoretical quantum mechanics, analogous classical mechanics, but in which only relations between observable quantities occur"} \cite{Heisenberg1925}. Only few scientists, first of all Einstein, realized at that time and later on that this proposal presupposes a revolutionary revision of the aim of science and even a new Weltanschauung. The essence of this revolutionary revision was expressed by Einstein in his explanation of {\it "reasons which keep he from falling in line with the opinion of almost all contemporary theoretical physicists"}: {\it "What does not satisfy me in that theory, from the standpoint of principle, is its attitude towards that which appears to me to be the programmatic aim of all physics: the complete description of any (individual) real situation (as it supposedly exists irrespective of any act of observation or substantiation)"} \cite{Einstein1949}. The philosophical fundamentals of QM, proclaimed by Heisenberg as far back as 1927 are: subjectivity, {\it "I believe that one can fruitfully formulate the origin of the classical 'orbit' in this way: the 'orbit' comes into being only when we observe it"} \cite{Heisenberg1927}; the negation of an objective reality, {\it "As the statistical character of quantum theory is so closely linked to the inexactness of all perceptions, one might be led to the presumption that behind the perceived statistical world there still hides a 'real' world in which causality holds. But such speculations seem to us, to say it explicitly, fruitless and senseless. Physics ought to describe only the correlation of observations"} \cite{Heisenberg1927}; indeterminism, {\it "One can express the true state of affairs better in this way: Because all experiments are subject to the laws of quantum mechanics, and therefore to equation (1), it follows that quantum mechanics establishes the final failure of causality"} \cite{Heisenberg1927}. The equation (1) in \cite{Heisenberg1927} is the famous Heisenberg's uncertainty relation. Thus, the uncertainty principle results to indeterminism, according to its author. 

\subsection{Quantum mechanics rejects the Cartesian polarity between 'res cogitans' and  'res extensa'}
Later on Heisenberg had developed his new Weltanschauung more neatly, in particular in his Lectures 1955-1956 "Physics and Philosophy" \cite{Heisenberg1959}. In the beginning of the Section V "The Development of Philosophical Ideas Since Descartes in Comparison with the New Situation in Quantum Theory" he stated: {\it "This reality was full of life and there was no good reason to stress the distinction between matter and mind or between body and soul"} \cite{Heisenberg1959}. This point of view by Heisenberg contradicts fundamentally to the scientific Weltanschauung of the previous centuries and Heisenberg emphasizes that QM compels to change this Weltanschauung. He reminds: {\it "The first great philosopher of this new period of science was Rene Descartes who lived in the first half of the seventeenth century. Those of his ideas that are most important for the development of scientific thinking are contained in his Discourse on Method"} \cite{Heisenberg1959}. And then Heisenberg points out on the importance of the Cartesian philosophy for the posterior development of natural science: {\it "While ancient Greek philosophy had tried to find order in the infinite variety of things and events by looking for some fundamental unifying principle, Descartes tries to establish the order through some fundamental division $\cdot \cdot \cdot $ If one uses the fundamental concepts of Descartes at all, it is essential that God is in the world and in the I and it is also essential that the I cannot be really separated from the world. Of course Descartes knew the undisputable necessity of the connection, but philosophy and natural science in the following period developed on the basis of the polarity between the 'res cogitans' and the 'res extensa', and natural science concentrated its interest on the 'res extensa'. The influence of the Cartesian division on human thought in the following centuries can hardly be overestimated, but it is just this division which we have to criticise later from the development of physics in our time"} \cite{Heisenberg1959}. 

The latter sentence clarifies in a greatest extent the fundamental difference of QM from all other theories of physics. All other theories concentrated their interest on the 'res extensa', i.e. all objects of the Nature, existing irrespective of any act of observation and the mind of the observer, i.e. the 'res cogitans'. Heisenberg points out on a philosophical basis of these theories {\it "Since $\cdot \cdot \cdot $ the 'res cogitans' and the 'res extensa' were taken as completely different in their essence, it did not seem possible that they could act upon each other"} \cite{Heisenberg1959}. He attacks this basis {\it "Obviously this whole description is somewhat artificial and shows the grave defects of the Cartesian partition"} but admits {\it "On the other hand in natural science the partition was for several centuries extremely successful. The mechanics of Newton and all the other parts of classical physics constructed after its model started from the assumption that one can describe the world without speaking about God or ourselves"} \cite{Heisenberg1959}. Before the QM emergence {\it "This possibility soon seemed almost a necessary condition for natural science in general. But at this point the situation changed to some extent through quantum theory"} \cite{Heisenberg1959}. Therefore Heisenberg comes {\it "to a comparison of Descartes's philosophical system with our present situation in modern physics"} \cite{Heisenberg1959}. 

His next remark demystifies the essence of his contradictions with Einstein and other critics of QM: {\it "If one follows the great difficulty which even eminent scientists like Einstein had in understanding and accepting the Copenhagen interpretation of quantum theory, one can trace the roots of this difficulty to the Cartesian partition. This partition has penetrated deeply into the human mind during the three centuries following Descartes and it will take a long time for it to be replaced by a really different attitude toward the problem of reality"} \cite{Heisenberg1959}. According to Heisenberg an old-fashioned attitude toward the problem of reality may be called dogmatic realism and metaphysical realism \cite{Heisenberg1959}. He explains the essence of the first: {\it "Dogmatic realism claims that there are no statements concerning the material world that cannot be objectivated $\cdot \cdot \cdot $ actually the position of classical physics is that of dogmatic realism. It is only through quantum theory that we have learned that exact science is possible without the basis of dogmatic realism. When Einstein has criticised quantum theory he has done so from the basis of dogmatic realism"} \cite{Heisenberg1959}. Einstein said {\it "I like to think that the moon is there even if I don't look at it"}, explaining his dislike for QM. Heisenberg and Einstein did not agree but they discussed on the common language of European philosophy. Therefore this controversy of they makes quite clear the essence of dogmatic realism. There is important also to know the essence of metaphysical realism according to Heisenberg: {\it "Metaphysical realism goes one step further than dogmatic realism by saying that 'the things really exist'. This is in fact what Descartes tried to prove by the argument that 'God cannot have deceived us'"} \cite{Heisenberg1959}. Heisenberg reminded above: {\it "On the basis of doubt and logical reasoning he} [Descartes] {\it tries to find a completely new and as he thinks solid ground for a philosophical system. He does not accept revelation as such a basis nor does he want to accept uncritically what is perceived by the senses. So he starts with his method of doubt. He casts his doubt upon that which our senses tell us about the results of our reasoning and finally he arrives at his famous sentence: 'cogito ergo sum'. I cannot doubt my existence since it follows from the fact that I am thinking. After establishing the existence of the {\bf I} in this way he proceeds to prove the existence of God essentially on the lines of scholastic philosophy. Finally the existence of the world follows from the fact that God had given me a strong inclination to believe in the existence of the world, and it is simply impossible that God should have deceived me"} \cite{Heisenberg1959}. Soon after Descartes his faith that God can not deceive was call in question by representatives for early empiristic philosophy, Locke, Berkeley and Hume. According to Heisenberg: {\it "The criticism of metaphysical realism which has been expressed in empiristic philosophy is certainly justified in so far as it is a warning against the naive use of the term 'existence'"} \cite{Heisenberg1959}. 

\subsection{The notion of the 'thing-in-itself' by Kant and hidden-variables }
According to the empiristic philosophy {\it "to be perceived is identical with existence"} \cite{Heisenberg1959}. Heisenberg admitted: {\it "This line of argument then was extended to an extreme scepticism by Hume, who denied induction and causation and thereby arrived at a conclusion which if taken seriously would destroy the basis of all empirical science"} \cite{Heisenberg1959} but he followed just this line when he rejected the 'thing-in-itself' and the law of causality of the Kant's philosophy. Heisenberg wrote: {\it "The disagreeable question whether 'the things really exist', which had given rise to empiristic philosophy, occurred also in Kant's system. But Kant has not followed the line of Berkeley and Hume, though that would have been logically consistent. He kept the notion of the 'thing-in-itself' as different from the percept, and in this way kept some connection with realism"} \cite{Heisenberg1959}. He was right that {\it "Considering the Kantian 'thing-in-itself' Kant had pointed out that we cannot conclude anything from the perception about the 'thing-in-itself'"} \cite{Heisenberg1959}. But his interpretation of the Kantian 'thing-in-itself' is very doubt: {\it "This statement has, as Weizsacker has noticed, its formal analogy in the fact that in spite of the use of the classical concepts in all the experiments a non-classical behaviour of the atomic objects is possible. The 'thing-in-itself' is for the atomic physicist, if he uses this concept at all, finally a mathematical structure: but this structure is - contrary to Kant - indirectly deduced from experience"} \cite{Heisenberg1959}.  

The Kantian 'thing-in-itself' is rather a cause of our perceptions than a mathematical structure. Any mathematical structure is a method of description and cannot belong to the percept or the perception. It can only describe they. Heisenberg and Weizsacker obscured the obvious meaning of the Kantian 'thing-in-itself' as a cause of our perceptions because of their persuasion that causality {\it "can have only a limited range of applicability"} \cite{Heisenberg1959}. They rejected the 'thing-in-itself' as the cause of our perceptions as well as they rejected hidden-variables as the cause of an individual observation of quantum phenomenon. 

\subsection{The Kantian a priori character of the law of causality and quantum mechanics }
The most doubt principle of QM and the new Weltanschauung by Heisenberg is indeterminism. Considering the law of causality Heisenberg wrote: {\it "Kant says that whenever we observe an event we assume that there is a foregoing event from which the other event must follow according to some rule. This is, as Kant states, the basis of all scientific work. In this discussion it is not important whether or not we can always find the foregoing event from which the other one followed. Actually we can find it in many cases. But even if we cannot, nothing can prevent us from asking what this foregoing event might have been and to look for it. Therefore, the law of causality is reduced to the method of scientific research; it is the condition which makes science possible. Since we actually apply this method, the law of causality is 'a priori' and is not derived from experience"} \cite{Heisenberg1959}. If the law of causality is the method of scientific research, as Kant stated, then QM proposed by Heisenberg is obviously no-scientific theory. Heisenberg tried to prove that {\it "the scientific method actually changed in this very fundamental question since Kant"} \cite{Heisenberg1959}. 

His first argument: {\it "We have been convinced by experience that the laws of quantum theory are correct and, if they are, we know that a foregoing event as cause for the emission at a given time cannot be found"} \cite{Heisenberg1959}. The other argument: {\it "We know the foregoing event, but not quite accurately. We know the forces in the atomic nucleus that are responsible for the emission of the $\alpha $-particle. But this knowledge contains the uncertainty which is brought about by the interaction between the nucleus and the rest of the world. If we wanted to know why the $\alpha $-particle was emitted at that particular time we would have to know the microscopic structure of the whole world including ourselves, and that is impossible"} \cite{Heisenberg1959}. Both arguments are doubt and have no relation to the fundamental question about the law of causality. If this law is 'a priori' and is not derived from experience then its disproof can not be also derived from experience. At least, the arguments by Heisenberg could not satisfy Schrodinger, proposing the cat paradox, Einstein and other critics of QM. Nevertheless Heisenberg stated that {\it "Kant's arguments for the a priori character of the law of causality no longer apply"} \cite{Heisenberg1959}. Most believers in the soothing philosophy or religion of Heisenberg-Bohr overlook this philosophical statement by Heisenberg. 

\section{Fundamental mistakes by sleepwalkers } 
Most sleepwalkers stride unimpeded, first of all, through the new Weltanschauung proposed by Heisenberg. Partly this carelessness may be explained with the inconsistency of Heisenberg. In the same paper \cite{Heisenberg1927} Heisenberg refutes a 'real' world in which causality holds and substantiates his uncertainty relation with help of the famous 'uncertainty microscope' existing in this real world in which causality holds. Later on he criticises the Cartesian polarity between the 'res cogitans' and the 'res extensa' and notes: {\it "In classical physics science started from the belief - or should one say from the illusion? - that we could describe the world or at least parts of the world without any reference to ourselves"} \cite{Heisenberg1959}. On the other hand he states that {\it "in the Copenhagen interpretation of quantum theory we can indeed proceed without mentioning ourselves as individuals"} \cite{Heisenberg1959}. Then, what is fundamental difference between the classical physics and the Copenhagen interpretation? Heisenberg confuses constantly the 'res cogitans' and the 'res extensa'. For example, disproving the claim by Alexandrov that {\it "Mention of the observer must be avoided"} (see above) Heisenberg writes: {\it "Of course the introduction of the observer must not be misunderstood to imply that some kind of subjective features are to brought into the description of nature. The observer has, rather, only the function of registering decisions, i.e. processes in space and time, and it does not matter whether the observer is an apparatus or a human being $\cdot \cdot \cdot $}" \cite{Heisenberg1959}. Any apparatus belongs to the 'res extensa' whereas any human being is the 'res cogitans', at least according to Descartes. If the observer is an apparatus then the Cartesian division should not be criticised {\it "from the development of physics in our time"} \cite{Heisenberg1959} and Heisenberg rather follows than disproves Alexandrov stating that {\it "'result of measurement' in quantum theory only the objective effect of the interaction of the electron with a suitable object"}, see above. The new Weltanschauung by Heisenberg could be unacceptable not only for most physicists but even for Heisenberg himself if it would be logically consistent.

\subsection{The quantum postulate and complementarity proposed by Bohr 'objectivate' observation} 
Heisenberg defines: {\it "The position to which the Cartesian partition has led with respect to the 'res extensa' was what one may call metaphysical realism. The world, i.e., the extended things, 'exist'"} \cite{Heisenberg1959}. According to this definition and the Cartesian division the dogmatic realism claims that there are no statements concerning the 'res extensa' that cannot be conceived outside of and independently of the 'res cogitans'. Therefore according to Heisenberg the term 'to objectivate' should signify to conceive any real situation outside of and independently of the mind of the observer. But he gives fundamentally different definition: {\it "We 'objectivate' a statement if we claim that its content does not depend on the conditions under which it can be verified"} \cite{Heisenberg1959}. Heisenberg as well as most theoretical physicists confuses here what is not measurable with the unreal. Therefore his practical realism is very vague. {\it "Practical realism assumes that there are statements that can be objectivated and that in fact the largest part of our experience in daily life consists of such statements"} \cite{Heisenberg1959}. If the term 'to objectivate' signifies only independence on the conditions of verification than the Heisenberg's practical realism can not make a distinction between the orthodox QM and theories of hidden variables. This distinction can be made only with help of the philosophically true definition of the term 'to objectivate'. We 'objectivate' a statement if we claim that a matter of its description (belonging to the 'res extensa') exists outside of and independently of the mind of the observer (belonging to the 'res cogitans'). According to this definition hidden variables can be objectivated whereas the uncertainty relation, the superposition of states, the Dirac jump and the act of observation in QM can not be objectivated. 

But the true definition of the term 'to objectivate' as well as the true essence of QM of Heisenberg-Bohr could be unacceptable for most physicists. Therefore QM based on the quantum postulate and complementarity by Bohr had became the symbol of almost general faith in spite of its self-contradiction. Bohr had objectivated the act of observation when he considered it as an interaction between quantum system and measuring instrument. It must be obvious that the Dirac jump can not be described by this way. Nevertheless most physicists including such eminent one as Feynman \cite{FeynmanL} and Landau \cite{LL} had followed Bohr in his wrong belief. Bell wrote that {\it "Landau sat at the feet of Bohr"} \cite{EPR1935}. Therefore Landau was sure {\it "that, in speaking of 'performing a measurement', we refer to the interaction of an electron with a classical 'apparatus', which in no way presupposes the presence of an external observer"} \cite{LL} and could not understand the logical absurdity of the LL jump. Even the EPR correlation \cite{EPR1935} could not shake the wrong belief of Bohr and his followers. It must be obvious that non-locality of the EPR correlation precludes any possibility to interpret the act of observation realistically as an interaction with measuring instrument. Nevertheless Bohr tried to save his quantum postulate and his complementarity. In his reply \cite{Bohr1935EPR} on the EPR paper \cite{EPR1935} Bohr criticizes the EPR criterion of the existence of an element of physical reality: {\it "$\cdot \cdot \cdot $ the wording of the above mentioned criterion $\cdot \cdot \cdot $ contains an ambiguity as regards the meaning of the expression 'without in any way disturbing a system'. Of course there is in a case like that just considered no question of a mechanical disturbance of the system under investigation during the last critical stage of the measuring procedure. But even at this stage there is essentially the question of an influence on the very conditions which define the possible types of predictions regarding the future behaviour of the system $\cdot \cdot \cdot $ their argumentation does not 'justify their conclusion that quantum mechanical description is essentially incomplete $\cdot \cdot \cdot $ This description may be characterized as a rational utilization of all possibilities of unambiguous interpretation of measurements, compatible with the finite and uncontrollable interaction between the objects and the measuring instruments in the field of quantum theory"}. 

This criticism is very obscure and Bell quoting it writes in \cite{Bell1981}: {\it "Indeed I have very little idea what this means. I do not understand in what sense the word 'mechanical' is used, in characterising the disturbances which Bohr does not contemplate, as distinct from those which he does. I do not know what the passage means - 'an influence on the very conditions $\cdot \cdot \cdot $'. Could it mean just that different experiments on the first system give different kinds of information about the second? But this was just one of the main points of EPR, who observed that one could learn either the position or the momentum of the second system. And then I do not understand the final reference to 'uncontrollable interactions between measuring instruments and objects', it seems just to ignore the essential point of EPR that in the absence of action at a distance, only the first system could be supposed disturbed by the first measurement and yet definite predictions become possible for the second system. Is Bohr just rejecting the premise - 'no action at a distance' - rather than refuting the argument?"} Indeed, Bohr could save his quantum postulate and his complementarity only rejecting the premise - 'no action at a distance'. {\it "The quantum postulate implies that any observation of atomic phenomena will involve an interaction with the agency of observation not to be neglected"} \cite{Bohr1928}. EPR had proposed a method of the observation, for example, by Alice the spin state of the distant particle flying toward Bob at which no interaction between the agency of observation of Alice and the Bob's particle can be assumed without action at a distance between they. The Bohr's complementarity can be valid also if only the premise - 'no action at a distance' could be rejected. This action should be real because of the reality of an interaction with the agency of observation which belongs to the 'res extensa', as well as any quantum object. The mind of Alice and Bob belong to the 'res cogitans'. But the quantum postulate and complementarity exclude an interaction with the mind of the observer. Bohr and his followers objectivated (in the meaning of the true definition) the act of observation, the Dirac jump and even the EPR correlation. The objectiveness of the EPR correlation implies a real action at a distance between the 'res extensa' contradicting to the relativity. The neglect by sleepwalkers this essential conflict {\it "between the two fundamental pillars of contemporary theory"} has resulted to mass delusion.

\subsection{The mass delusion and the idea of quantum computation}
This mass delusion shows itself, in particular, in the idea of quantum computation enjoying wide popularity now \cite{Nielsen2000}. The widespread interest in this idea seems quite valid. The minimal sizes of nanostructures come nearer to atomic level and subsequent miniaturization will not be possible in the near future. Therefore exponential increase of calculating resources with number of quantum bits has provoked almost boundless enthusiasm. This exponential increase seems possible thanks to the principle of superposition of states, interpreted as the cardinal positive principle of the QM \cite{LL}. Feynman \cite{Feynman1982} proposed universal simulation, i.e. a purpose-built quantum system which could simulate the physical behaviour of any other, noting that the calculation complexity of quantum system increases exponentially with number $N$ of its elements. Indeed, the number $g_{N} =2^{N} - 1$ of independent variables $\gamma _{j}$ describing, for example, the spin 1/2 states of $N$ particle
$$\psi  = \gamma _{1}|\uparrow  \uparrow \uparrow ...\uparrow > + \gamma _{2}|\uparrow  \uparrow \uparrow ... \downarrow >+ ... $$ $$ +  \gamma _{gN-1}|\downarrow  \downarrow \downarrow... \uparrow > +  \gamma _{gN}|\downarrow  \downarrow \downarrow... \downarrow > \eqno{(15)}$$
increases exponentially with the number of these particles thanks to the EPR correlation. 

The Feynman's idea of simulation was based on his belief in QM as an universal theory. But this belief is false. The idea of universal quantum computer  was proposed by David Deutsch {\it "as a way to experimentally test the 'Many Universes Theory' of quantum physics - the idea that when a particle changes, it changes into all possible forms, across multiple universes"} \cite{FatherQC2007}. There is important to remind that the concept of multiple universes was proposed by Hugh Everett \cite{Everett1957} in order to describe the Process 1, i.e. the act of observation out the domain of psychology, i.e. without the mind of the observer. In the early 1990's several authors sought computational tasks which could be solved by a quantum computer more efficiently than any classical computer. Shor has described in 1994 \cite{Shor1994} an algorithm which was not only efficient on a quantum computer, but also addressed a central problem in computer science: that of factorising large integers. This possibility has provoked mass enthusiasm. But most authors of numerous publication on quantum computation ignore the statement of the author of this idea that quantum computer can be real only in multiple universes. Deutsch writes in his book \cite{Deutsch1997FR}: {\it "For whose who still thinks that there is only one universe I offer the following problem: to explain a principle of action of the Shor's algorithm. I do not request to predict that it will work, as for this purpose it is enough to solve some consistent equations. I ask you to give an explanation. When the Shor's algorithm has factorized number, having involved about $10^{500}$ computing resources which can be seen where this number was factorized on multipliers? In whole visible universe exists in all about $10^{80}$ atoms, the number is insignificant small in comparison with $10^{500}$. Thus, if the visible universe was a whole physical reality, the physical reality even is remote would not contain resources, sufficient for factorization on multipliers of such big number"}. 

This contradiction between the author of the idea of quantum computation and numerous authors of publications about quantum computation is consequence of the robust common sense intrinsic not only the authors \cite{Kampen1988} but almost all physicists. Deutsch, as well as Everett, understands that the problem of 'observation' in the orthodox QM {\it "cannot be ruled out as lying in the domain of psychology"} \cite{Everett1957} and only the idea of multiple universes can deliver from this nonsense. In contrast to Deutsch and Everett most authors believing Heisenberg and Bohr, who had obscured the logically obvious fact, spurn such mind-boggling fantasies as both the many world interpretation and the mind of the observer having an influence on quantum system at observation. The belief in the soothing philosophy or religion of Heisenberg-Bohr is so thoughtless that believers refuse to admit that the numerous independent variables $\gamma _{j}$ in the EPR correlation (15) describe the knowledge of an observer. It must be obvious from the fact that the huge number $g_{N} =2^{N} - 1$ decreases twice, for example approximately on $10^{1000}$ at the number of quantum bits $N \approx  2303$, when the observer observes the spine state only one particle.  

Our mind, in a certain relation, is much more rich than our empirical knowledge. We can write any big number, for example $10^{1000}$, $10^{10000000}$, $10^{1000000000}$, and even to set yourself a task to factorized it on multipliers with help of the Shor's algorithm. But it does not mean, that this task can be solved with help of a real device, quantum computer, in a reality of one universe. It could be possible only if a real action on distance, violating relativistic causality, could be possible. The violations \cite{Aspect1981,Aspect1982,Aspect82} of the Bell's inequalities give experimental evidence of {\it "a gross violation of relativistic causality"}, p. 171 in \cite{Bell2004}, for results of observations. The EPR correlation describes just this violation. But the experimental results \cite{Aspect1981,Aspect1982,Aspect82} and any others can not give evidence that the EPR correlation describes a real action on distance between the 'res extensa', existing outside of and independently of the 'res cogitans', i.e. the mind of the observer. Authors of numerous publications about quantum computation objectivate without hesitation quantum bits and quantum gates doing not suspect about conflict with relativistic causality and even with the Copenhagen interpretation of quantum theory. In the Section "The Copenhagen Interpretation of Quantum Theory" of \cite{Heisenberg1959} Heisenberg stated {\it "that certainly our knowledge can change suddenly and that this fact justifies the use of the term 'quantum jump'"}. Just this sudden change of our knowledge at observation results to non-locality of the EPR correlation and the conflict between the Copenhagen interpretation of quantum theory with relativistic causality. Heisenberg noted also in this Section that {\it "there is no description of what happens to the system between the initial observation and the next measurement"} \cite{Heisenberg1959}. Thus, according to the Copenhagen interpretation QM can not describe a process of quantum computation which should be just before the next measurement. Both the EPR correlation and quantum computation can not be possible according to hidden-variables theories. Deutsch is quite right that quantum computer can be real only in a reality of multiple universes. 

\subsection{Two principal mistakes of Heisenberg}
The Heisenberg's point of view was inconsistent. On the one hand he understood that {\it "if a interaction of a system with the measuring apparatus is treated as a whole according to quantum mechanics and if both are regarded as cut off from the rest of the world, then the formalism of quantum theory does not as a rule lead to a define result"} \cite{Heisenberg1959}. But on the other hand he did not want to admit the presence of subjectivity in his pet quantum mechanics. Although the subjectivity follows logically and inevitably from his understanding that 'quantum jump' of the measuring apparatus can not be spontaneous. Trying to deny subjective features of QM Heisenberg was forced to use obscure declarations, for example: {\it "The observer has, rather, only the function of registering decisions, i.e. processes in space and time, and it does not matter whether the observer is an apparatus or a human being; but the registration, i.e. the transition from the 'possible' to the 'actual' is absolutely necessary here and can not be omitted from the interpretation of quantum theory. At this point quantum theory is intrinsically connected with thermodynamics in so far as every act of observation is by its very nature an irreversible process; it is only through such irreversible processes that the formalism of quantum theory can be consistently connected with actual events in space and time"} \cite{Heisenberg1959}. This explanation is very obscure! Any apparatus belongs to the 'res extensa' whereas a human being belongs to the 'res cogitans'. An answer on the question "What or who is the observer, the 'res extensa' or the 'res cogitans'?" should be decisive for the essence of QM. But Heisenberg ignores this question because he understands that no apparatus can induce the transition from the 'possible' to the 'actual'. This transition can be only in the mind of a human being. But Heisenberg did not want to admit this subjective feature of QM. 

The proclaim resemblance of QM and thermodynamics distracts the attention from the principal problem and has misled many physicists, in particular Landau \cite{LL}. The principal problem was raised by Bell in his paper "On the problem of hidden variables in quantum mechanics" \cite{Bell1966}: {\it "To know the quantum mechanical state of a system implies, in general, only statistical restrictions on the results of measurements. It seems interesting to ask if this statistical element be thought of as arising, as in classical statistical mechanics, because the states in question are averages over better defined states for which individually the results would be quite determined"}. Bell in this work had given the answer of the principal question "What or who is the observer?" If all quantum phenomena can be described using better defined states for which individually the results would be quite determined, i.e. 'hidden variables' than the observer is an apparatus. But in this case the orthodox QM proposed by Heisenberg is observably inadequate. It may be adequate if {\it "atomic and subatomic particles do not have any definite properties in advance of observation"} \cite{Bell1981}. In this case the observer must be the mind of a human being because no apparatus can create definite properties when they do not exist in advance of observation. Any apparatus can only change properties and make they hidden. 

The vagueness of QM is a consequence of logical mistakes made by young Heisenberg. Einstein warned against one of they. Heisenberg remembered \cite{Heisenberg1969} that after his talk at the Berlin university in 1926 Einstein has told him that {\it "from the fundamental point of view the intention to create a theory only on observable parameters completely ridiculously. Because in the real all is quite the contrary. Only the theory can decide what exactly can be observable. As you can see, observation, generally speaking, is very complicated process"}. In 63 years later Bell wrote: {\it "Einstein said that it is theory which decides what is 'observable'. I think he was right - 'observation' is a complicated and theory-laden business. Then that notion should not appear in the formulation of fundamental theory"} \cite{EPR1935}. Einstein and Bell asserted fairly that the offer by Heisenberg to describe only observable parameters rather has complicated than has simplified the task of theory. The description of observation results can not be complete without a description of the act of observation. Therefore it is easier to create a theory describing {\it "real situation (as it supposedly exists irrespective of any act of observation or substantiation)"} \cite{Einstein1949} than a theory describing observation results. The vagueness of the notion about observation (measurement) is principal cause of the fundamental obscurity in quantum. Bell was right when he stated that {\it "On this list of bad words from good books, the worst of all is 'measurement'"}. Heisenberg proposing to describe observation results could not even formulate enough clear what is observation. 

This logical mistake is result of other fundamental mistake concerning his conception of human thinking. Heisenberg noted: {\it "This basis of the philosophy of Descartes is radically different from that of the ancient Greek philosophers. Here the starting point is not a fundamental principle or substance, but the attempt of a fundamental knowledge. And Descartes realises that what we know about our mind is more certain than what we know about the outer world"} \cite{Heisenberg1959}. But Heisenberg himself seems to do not realise that our knowledge about our mind is more certain than about the outer world. Only therefore he could contest the logical conclusion by Kant that the law of causality is a priori form of our mind and the basis of all scientific work. The vagueness and self-contradiction of QM prove that rather Kant than Heisenberg was right. Considering the law of causality Heisenberg wrote: {\it "Kant says that whenever we observe an event we assume that there is a foregoing event from which the other event must follow according to some rule"} \cite{Heisenberg1959}. This conclusion by Kant about human thinking is confirmed with an interesting discussion between creators of quantum mechanics at the Solvay meeting 1927, which Bohr remembered in 1949 \cite{Bohr1949}: {\it "On that occasion an interesting discussion arose also about how to speak of the appearance of phenomena for which only predictions of statistical character can be made. The question was whether, as to the occurrence of individual effects, we should adopt a terminology proposed by Dirac, that we were concerned with a choice on the part of 'nature' or, as suggested by Heisenberg, we should say that we have to do with a choice on the part of the "observer" constructing the measuring instruments and reading their recording. Any such terminology would, however, appear dubious since, on the one hand, it is hardly reasonable to endow nature with volition in the ordinary sense, while, on the other hand, it is certainly not possible for the observer to influence the events which may appear under the conditions he has arranged"}. 

Here three principal creators of QM discuss what is the cause of the appearance of phenomena. None of they state the absence of the cause, even Heisenberg. Almost 30 years later Heisenberg calls in question that the Kant's conclusion {\it "Since we actually apply this method, the law of causality is 'a priori' and is not derived from experience"} \cite{Heisenberg1959} may be true in atomic physics. Heisenberg considers a radium atom, which can emit an $\alpha $-particle states {\it "The time for the emission of the $\alpha $-particle cannot be predicted"} \cite{Heisenberg1959}. Next he asks {\it "But why has the scientific method actually changed in this very fundamental question since Kant?"}, gives two answers one of they is {\it "We have been convinced by experience that the laws of quantum theory are correct and, if they are, we know that a foregoing event as cause for the emission at a given time cannot be found"} and concludes {\it "Therefore, Kant's arguments for the a priori character of the law of causality no longer apply"} \cite{Heisenberg1959}. This Heisenberg's conclusion is absolutely ill-founded because he considers the 'res extensa' whereas Kant considered the 'res cogitans', human mind. The Kant's conclusion that the law of causality is 'a priori' is 'a priori'. Therefore neither experience nor theory describing correctly this experience can call this conclusion in question. The negation of a cause in 'nature', i.e. the terminology proposed by Dirac results inevitably to the suggestion by young Heisenberg that the 'observer' is the cause of the occurrence of individual effects. Just this suggestion results to the EPR paradox, the Schrodinger's cat paradox, the questions "Whose knowledge and whose will?", considered above, and other nonsense.

\section{Fundamental mistakes because of the prejudice of the QM universality }
The fundamental obscurity in QM discussed during almost ninety years is connected with the vague notion about the act of observation, i.e. the Process 1 according to Everett \cite{Everett1957}. The description of the Process 1 should lie in the domain of psychology \cite{Everett1957}. Therefore it is very important to draw reader's attention that the act of observation is absent at the description of most quantum phenomena. It was noted in the Introduction that already Feynman had become aware of the applicability of the realistic interpretation of wave function  proposed by Schrodinger for description of macroscopic quantum phenomena. But he did not understand the fundamental importance of this fact. Anyone can easy see that not only macroscopic quantum phenomena are described with the Schrodinger interpretation, i.e. without the Process 1, the mind of the observer and other mysticism. Schrodinger had introduced the term 'wave function' for his interpretation and used, as well as Einstein \cite{Einstein1949}, the term '$\psi $ - function' \cite{Schrod35D} for the positivistic interpretation proposed by Born. The terms 'wave function' $\Psi _{Sh}$  and '$\psi $ - function' $\Psi _{B}$ will be used below in the same sense. It would be useful to categorise the quantum phenomena described the 'wave function' and the '$\psi $ - function'. QM uses the '$\psi $ - function' when the Dirac jump is used at the description and the 'wave function' when the Dirac jump is absent.

\subsection{The Aharonov - Bohm effects are described both with the $\psi $ - functions and the wave function }
For example, the Dirac jump is used for the description of the two-slit interference experiment demonstrating particle - wave duality at observation. The interference pattern, for example electrons observed on a detecting screen \cite{electronInt}, testifies to that electrons pass through the double-slit as a wave with the de Broglie wavelength $\lambda _{deB} = 2\pi \hbar /mv$. But the same electrons manifest itself as particles at they arrival at the detecting screen \cite{electronInt}, see also Fig.1-3 in the book \cite{QuCh2006}. QM describes this observed duality with help of the $\psi $ function $\Psi_{B} = \Psi _{B1} + \Psi _{B2}$ which is the superposition of two $\psi $ - functions $\Psi _{B1} = A_{1}e^{i\varphi _{1}}$, $\Psi _{B2}= A_{2}e^{i\varphi _{2}}$ describing two possible path $L_{1}$, $L_{2}$ of electrons through the first and second slits with the amplitudes $A_{1}$ and $ A_{2}$ of the arrival probability at the point $y$ of a particle passing through the first and second slit. Then, the probability $P(y) = |\Psi_{B}|^{2}$ of the electron arrival at a point $y$ of the detecting screen, Fig.3,   
$$P(y) = A_{1}^{2} + A_{2}^{2} + 2A_{1}A_{2} \cos(\Delta \varphi _{1} - \Delta \varphi _{2}) \eqno{(16)} $$
should depend on the phase difference $\Delta \varphi _{1} - \Delta \varphi _{2}$. Y. Aharonov and D. Bohm had noted more than fifty years ago \cite{AB1959} that this phase difference $\Delta \varphi _{1} - \Delta \varphi _{2} = \int_{S}^{y}dr_{1}(p/\hbar ) - \int_{S}^{y}dr_{2}(p/\hbar ) = \int_{S}^{y}dr_{1}(mv/\hbar ) - \int_{S}^{y}dr_{1}(mv/\hbar ) + \oint dr (eA/\hbar ) =2\pi  (L_{1} - L_{2})/\lambda _{deB} + 2\pi \Phi /\Phi _{0}$ and consequently the probability (16) should change with magnetic flux $\Phi $ inside the closed contour $S - L _{1} - y - L _{2} -S$, Fig.1, because of the relation $p = mv + qA$ between canonical momentum $p$ and electron $q = e$ velocity $v$ in the presence of a magnetic vector potential $A$.  The interference pattern shift with the flux $\Phi $ change was corroborated by many experiments \cite{ABRMP1985}. The value of the probability (16) oscillates with the period equal the flux quantum $\Phi _{0} = 2\pi \hbar /q$. 

\begin{figure}
\includegraphics{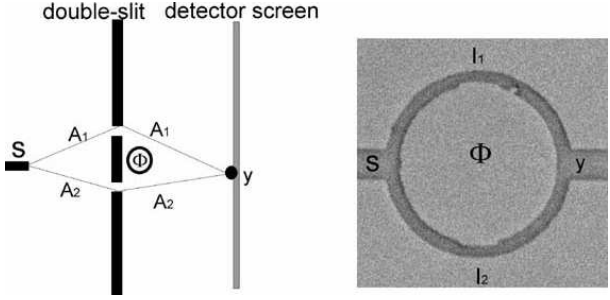}
\caption{\label{fig:epsart} The Ahronov-Bohm effects in the two-slit interference experiment (on the left) and in in normal metal or superconductor ring (on the right) are consequence of the magnetic flux $\Phi $ influence on the phase difference $\Delta \varphi _{1} - \Delta \varphi _{2} = \oint_{l}dl \nabla \varphi $. In the first case the difference $\Delta \varphi _{1} - \Delta \varphi _{2} $ of the phase changes between $S$ and $y$ points along upper $L _{1}$ way $\Delta \varphi _{1}$ and lower $L _{2}$ way $\Delta \varphi _{2}$ should not be divisible by $2\pi $ ($\Delta \varphi _{1} - \Delta \varphi _{2} \neq n2\pi $) since the $\psi $ function collapses at the observation of the electron arrival in a point $y$ of the detector screen. In contrast to the interference experiment the Ahronov-Bohm effects in mesoscopic ring (for instance the persistent current) are observed because of the requirement $\Delta \varphi _{1} - \Delta \varphi _{2} = n2\pi $ since the wave function, describing this case, can not collapse}
\end{figure} 

The periodicity because of the Aharonov - Bohm effect \cite{ABRMP1985}, i.e. because of the relation $\hbar \nabla \varphi = p = mv + qA$, are observed also in normal metal \cite{PCScien09} and superconductor \cite{PCScien07,PCJETP07,ABNanoSt2009} loops. But this effect differs in essence from the case of the two-slit interference experiment. The periodical dependencies in magnetic flux $\Phi $ with period $\Phi _{0} = 2\pi \hbar /q$ are observed in the mesoscopic rings \cite{PCScien09,PCScien07,PCJETP07,ABNanoSt2009} because of the quantization of velocity circulation 
$$\oint_{l}dl v  =  \frac{2\pi \hbar }{m}(n - \frac{\Phi}{\Phi_{0}})  \eqno{(17)}$$
The quantization (17) is deduced from the requirement that the complex wave function must be single-valued $\Psi _{Sh} = |\Psi _{Sh}|\exp i\varphi = |\Psi _{Sh}|\exp i(\varphi + n2\pi)$ at any point. Because of this requirement, the phase $\varphi $ must change by integral $n$ multiples of $2\pi $ following a complete turn along the path of integration, yielding $\oint_{l}dl \bigtriangledown \varphi = \oint_{l}dl p/\hbar = \oint_{l}dl (mv + qA)/ \hbar = m\oint_{l}dl v/ \hbar + 2\pi \Phi/\Phi_{0}= n2\pi $.

This requirement is violated at the description of the two-slit interference experiment because of the collapse of the $\psi $ - function (the Dirac jump) at the observation of the electron arrival in a point $y$ of the detector screen. The phase difference $\Delta \varphi _{1} - \Delta \varphi _{2} = \int_{S}^{y}dr_{1} \nabla \varphi - \int_{S}^{y}dr_{2} \nabla \varphi = \oint_{l} dr \nabla \varphi $ and the probability (16) can change uninterruptedly with the coordinate $y$ and magnetic flux $\Phi $ thanks to the collapse. This uninterrupted variation provides the interference pattern (16) and its shift with $\Phi $. Thus, although the both Aharonov - Bohm effects result from the $\Phi $ influence on the phase variation along a closed path $\oint_{l} dr \nabla \varphi $ they differ fundamentally: the first one should be described with the $\psi $ - function whereas the second one should be described with the wave function \cite{FPP2008}. 

The ignorance about this fundamental difference provokes mistakes. For example, the authors \cite{Strambini} have concluded that electrons can be reflected because of magnetic flux $\Phi $ in the Aharonov-Bohm ring, in defiance of the law of momentum conservation \cite{Comment2010}. The contradiction in such scandalous form is absent in the Aharonov - Bohm effect \cite{AB1959}, although there is a problem with non-local force free momentum transfer \cite{QuCh2006,Nature08}. In spite of the change in the interference pattern no {\it overall} deflection of electrons is observed in the Aharonov-Bohm effect because of magnetic flux \cite{QuCh2006}. The transmission (reflection) probability $P_{tr} = \int dy P(y) = \int dx (A_{1}^{2} + A_{2}^{2}) = 1$ can not depend at all on magnetic flux $\Phi $ contrary to the erroneous theoretical result shown on Fig.2 in \cite{Strambini}. The mistake \cite{Strambini} is one of consequences of the prejudice of the QM universality \cite{WakeUp}.

\subsection{We can believe for the time being in reality of the moon}
Even macroscopic realism was called in question because of this prejudice. Bell noted that creators of QM {\it "despair of finding any consistent space-time picture of what goes on the atomic and subatomic scale"} \cite{Bell1981}. {\it "For example \cite{Jammer74}, Bohr once declared when asked whether the quantum mechanical algorithm could be considered as somehow mirroring an underlying quantum reality: 'There is no quantum world. There is only an abstract quantum mechanical description. It is wrong to think that the task of physics is to find out how Nature is. Physics concerns what we can say about Nature'"} \cite{Bell1981}. The creators of QM disclaimed reality only on the atomic and subatomic scale. Recently this scepticism about realism was expanded to macroscopic level  \cite{Mooij2010}. The authors \cite{Mooij2010} quoting the known remark by Albert Einstein {\it "I like to think that the moon is there even if I don't look at it"} claims that some experimental results obtained on superconducting circuit \cite{Korotkov2010}  could refute this trust by Einstein in objective reality even on the macroscopic level. This scandalized claim bears a direct relation to the problem of a possibility of superconducting quantum bits \cite{Clarke2008}. A small moon, which is not there according to \cite{Mooij2010}, is a magnetic flux inside a superconducting loop \cite{Leggett1985} which is considered as flux qubit in numerous publications \cite{Clarke2008} including the one \cite{Mooij2003}  of the author of the scandalized claim \cite{Mooij2010}. The authors \cite{Mooij2010} believes paradoxically that QM can prove that nothing is there but it is possible to create anything, superconducting quantum bits \cite{Mooij2003} for example, using this nothing. Ironically, quantum bits can be created indeed only if quantum systems do not have any definite properties in advance of observation. 

Only basis both of the doubt in macroscopic reality \cite{Mooij2010} and of the reality of superconducting quantum bits \cite{Clarke2008,Mooij2003} is the Leggett-Garg inequality referred to as a Bell's inequality in time \cite{Mooij2010,Korotkov2010}. A.J. Leggett and A. Garg \cite{Leggett1985} have concocted the contradiction with realism extremely unsuccessful. The doubt about reality of macroscopic magnetic flux in rf SQUID, i.e. a superconducting loop interrupted by Josephson junction, is provoked in \cite{Leggett1985} only because of  $\psi $ - function usage in additional to the wave function describing superconducting state. $\psi $ - function can apply speculatively for description of any macroscopic object, for example the cat \cite{Schrod35D} or the moon \cite{Mermin1985}. But realism must not be called in question without irrefutable empirical evidence obtained on the base of no-hidden-variables theorem (or, vulgarly, 'no-go theorem') \cite{Mermin1993}.

The superconducting loop interrupted by Josephson junction, considered in \cite{Leggett1985}, has two permitted states with the same minimal energy and opposite directed superconducting current (the persistent current $I _{p}$), when external magnetic flux inside its loop is divisible by half $\Phi = (n+0.5)\Phi _{0}$ of the flux quantum $\Phi _{0} = 2\pi \hbar /q$, see the relation (17). Imitating Bell, A.J. Leggett and A. Garg \cite{Leggett1985} have offered inequalities which experimental check should prove the contradiction with any realistic theory and the necessity of the positivistic description with help of superposition of states, which should collapse at observation. The no-go theorem by A.J. Leggett and A. Garg \cite{Leggett1985} is false because of some reasons.  First of all there is not a well-grounded motive to call macroscopic realism in question because rather the realistic Schrodinger's than positivistic Born's interpretation of the wave function is valid for description of macroscopic quantum phenomena. The authors \cite{Leggett1985} as well many others \cite{Clarke2008,Mooij2003,Leggett2002} liken superconducting loop to spin 1/2 and assume superposition of its two permitted states, described as the superposition of the spin states $\psi  = \alpha  |\uparrow  > +  \beta  |\downarrow  >$. This likening is obviously false because of some reasons. First of all because superconducting loop is flat and the magnetic moment $M_{m} = I_{p}S$ of the current $I_{p}$ circulating in it anticlockwise or clockwise and the angular momentum of Cooper pairs $M_{p} = (2m_{e}/e)I_{p}S$ are one-dimensional. The assumption \cite{Leggett1985,Clarke2008,Mooij2003,Leggett2002} of superposition of states with different value of the one-dimensional angular momentum contradicts to formalism of QM \cite{LL} even if  the one-dimensional angular momentum is microscopic. Therefore it is very strange that A.J. Leggett and A. Garg \cite{Leggett1985} and other authors \cite{Clarke2008,Mooij2003,Leggett2002} could assume superposition of states with macroscopically different $\Delta M_{p} \approx  10^{5} \ \hbar $ angular momentum. This assumption contradicts obviously to the fundamental law of angular momentum conservation \cite{Nikulov2010} and is much more inconceivable than superposition of the cat's states, at least of the Schrodinger's cat (9). Besides the authors \cite{Leggett1985} repeat virtually the mistake \cite{Bell1966,Mermin1993} of the von Neumann's no-hidden-variables proof. Thus, the doubt in macroscopic realism \cite{Leggett1985,Mooij2010} is absolutely baseless and was provoked with misinterpretation of QM as a theory describing universally all quantum phenomena. We can believe as before in reality of the moon.

\section{Fundamental obscurity connected with wave function usage}
The wave function describes a {\it real situation (as it supposedly exists irrespective of any act of observation or substantiation)} in accordance with {\it the programmatic aim of all physics} upheld by Einstein \cite{Einstein1949}. Therefore the fundamental obscurity connected with the Born's interpretation is absent at the description of macroscopic and some other quantum phenomena. But there is different fundamental obscurity, more real than the one connected with violation of the Bell's inequalities. In spite of this obscurity, the universally recognised quantum formalism describes almost all effects observed in superconductor structures. But some experimental results obtained recently contradict to its predictions.

\subsection{Puzzles generated with the quantum formalism}
The real situation, described with the wave function $\Psi _{Sh} = |\Psi _{Sh}|\exp i\varphi $, can not change because of our look, but it can alter with help of a real physical influence. For example, we can decrease the density of Cooper pairs $|\Psi _{Sh}|^{2} = n_{s} = n_{s,0}(1 - T/T_{c})$ down to $n_{s} = 0$ at a time $t = t_{on}$, overheating a loop segment $B$ above superconducting transition $T > T_{c}$, for example with help of laser beam, Fig.4. The electric field $E = - \bigtriangledown V \approx V_{B}/(l-l_{B})$ of the potential voltage 
$$V_{B} = R_{B}I(t)= R_{B}I_{p}\exp -\frac{t - t_{on}}{\tau_{RL}} \eqno{(18)} $$ 
appeared because of a non-zero resistance $R_{B} > 0$ of the $B$ segment in normal state, will decrease the velocity of Cooper pairs from the quantum value (17), equal $v = -2\pi \hbar/lm4$ at the magnetic flux inside the loop $\Phi = \Phi_{0}/4$, down to zero $v = 0$, during the relaxation time $\tau_{RL} = L/R_{B}$. $I_{p} = s2en_{s}v$ is the persistent current observed because of the quantization (17); $s = hw$ is the loop section; $L$ is the inductance of the loop with a length $l$. This velocity change occurs in accordance with the Newton's second law $mdv/dt = 2eE$, under the influence of the real force $F_{E} = 2eE$ acting on each Cooper pair. According to the universally recognised quantum formalism the velocity in all loop segment, including the $A$ one, Fig.4, must return to the initial value $v = -2\pi \hbar/lm4$ corresponding the permitted state (17) with minimum energy $\propto (n - \Phi /\Phi_{0})^{2}$ after turning off of the laser beam at a time $t_{off}$ and the cooling of the $B$ segment down to the initial temperature $T < T_{c}$, see the left picture on Fig.4. The pair velocity in the segment $A$ should change without a real force at the change of the real situation in the spatially separated segment $B$, Fig.4, because of the prohibition (17) of the zero velocity $v = 0$ at $\Phi \neq n\Phi_{0}$. The quantum formalism can not explain this non-local force-free momentum transfer. It can only describe phenomena connected with it. 

\begin{figure}
\includegraphics{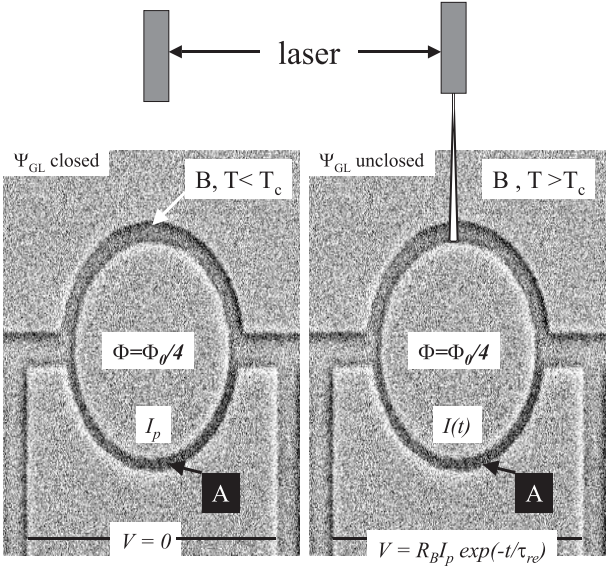}
\caption{Superconducting loop can be switched between the states with different connectivity of the wave function $\Psi _{Sh} = |\Psi _{Sh}|\exp i\varphi $ with a real physical influence, for example turning on (the right picture) and turning off (the left picture) of the laser beam heating the loop segment $B$ above $T_{c}$. The persistent current, equal $I_{p} = -s2en_{s}(2\pi \hbar/lm4)$ in a symmetric loop $l$ with the same section $s$ and pair density $n_{s}$ along the whole  $l$, flows at the magnetic flux $\Phi = \Phi_{0}/4$ inside $l$ when the wave function is closed (the left picture). The current circulating in the loop should decay during the relaxation time $\tau_{RL} = L/R_{B}$ after the transition in the state with unclosed wave function because of a non-zero resistance $R_{B} > 0$ of the $B$ segment in the normal state (the right picture). The photo of a real aluminum loop is used in order to exhibit that the gedankenexperiment can be made real.}
\label{fig:1}       
\end{figure}

The direct component 
$$V_{dc} = \frac{1}{\Theta }\int_{\Theta }dtV_{B}(t) \approx  L\omega _{sw} \overline{I_{p}}; \  at \ \omega _{sw}\tau_{RL} \ll 1  \eqno{(19a)}$$
$$  V_{dc} \approx  R_{B} \overline{I_{p}}; \ at \ \omega _{sw}\tau_{RL} \gg 1 \eqno{(19b)}$$
of the voltage (12) should be observed at repeated switching with a frequency $\omega _{sw} $ of the $B$ segment, Fig.4, between superconducting and normal states, because of the returning of the loop to the same state $n$ at each $B$ cooling \cite{JLTP1998}. The sign and value of the dc voltage (19) should vary periodically with magnetic flux $V_{dc}(\Phi /\Phi_{0})$ like the persistent current $I_{p}(\Phi /\Phi_{0})$ \cite{PCScien07}, because of the change of the quantum number $n$ corresponding to minimum energy $\propto (n - \Phi /\Phi_{0})^{2}$ at $\Phi = (n+0.5)\Phi_{0}$ \cite{Tinkham}. Such quantum oscillations of the dc voltage $V_{dc}(\Phi /\Phi_{0}) \propto I_{p}(\Phi /\Phi_{0})$ were observed on segments of asymmetric aluminium rings when the switching take place because of noise \cite{PhysLetA2012,toKulik2010,Letter07,PerMob2001} or ac current \cite{PCJETP07,Letter2003}. The experimental results \cite{Letter07,toKulik2010} give unequivocal evidence that the persistent current can flow against the dc electric field $E = -\bigtriangledown V$. This puzzle may be connected with the other one: the observations \cite{PCJETP07,ABNanoSt2009,PCScien07,toKulik2010,Letter07,LP1962} of the persistent current $I_{p}(\Phi /\Phi_{0})$ at a non-zero resistance $R_{l} > 0$. An electric current should decay during a very short relaxation time $\tau_{RL} = L/R_{l} < 10^{-9} \ s $ without Faraday's voltage $ -d\Phi /dt = 0$ in the aluminium ring, used in \cite{Letter07,toKulik2010,PCScien07}, with radius $r \approx 1 \ \mu m$, the inductance $L \approx 10^{-11} \ H$, at resistance $R_{l} > 0.01 \ \Omega $. In defiance of this the persistent current does not decay \cite{Letter07,toKulik2010,PCScien07}.

This puzzles can be described in the limits of the quantum formalism taking into account that the angular momentum $\oint_{l}dl p = \oint_{l}dl (mv + 2eA) = m\oint_{l}dl v + 2e\Phi $ of each Cooper pair should change from $2e\Phi $ to $n2\pi \hbar $ because of the quantization at each closing of superconducting state in the ring. This change of pair momentum $p$ in a time unit at repeated switching with a frequency $\omega _{sw} $ was called in \cite{PRB2001} "quantum force" $F_{q}$: 
$$\oint_{l}dlF_{q}=2\pi \hbar (\overline{n}- \frac{\Phi }{\Phi_{0}})\omega _{sw}  \eqno{(20)}$$
at $\omega _{sw} \ll 1/\tau_{RL}$. The quantum force $\oint_{l}dlF_{q}$ takes the place of the Faraday's voltage $-d\Phi /dt$ which maintains $IR_{l} = -d\Phi /dt$ a conventional current $I$ circulating in a loop and describes why the persistent current can not decay $\overline{I_{p}}R_{l} = \oint_{l}dlF_{q}/2e$ in spite of the power dissipation $\overline{I_{p}^{2}R_{l}}$. Under equilibrium condition the $I_{p} \neq 0 $ at $R_{l} > 0$ is observed only in a narrow temperature region near superconducting transition $T \approx T_{c}$ \cite{Letter07,toKulik2010,PCScien07}, where thermal fluctuations can switch loop segments between superconducting $n_{s} > 0$, $R = 0$ and normal $n_{s} = 0$, $R > 0$ states \cite{PRB2001}.

The first experimental evidence of the force-free angular momentum transfer was obtained as far back as 1933 when Meissner and Ochsenfeld \cite{Meissner1933} observed first that a superconductor, placed in a weak magnetic field $B < B_{c1}$, completely expels the field from the superconducting material except for a thin layer $\lambda _{L} \approx 50 \ nm = 5 \ 10^{-8} \ m$ at the surface. The quantum formalism describes the Meissner effect as the particular case $n = 0$ of the flux quantization $\Phi = n\Phi_{0} = 0$ \cite{FPP2008} but can not explain the puzzle. In the case, considered on Fig.4, the angular momentum change of single pair $ n2\pi \hbar - 2e\Phi = 2\pi \hbar (n - \Phi /\Phi_{0}) \leq 2\pi \hbar  \ 0.5$ is microscopic irrespective of the loop $l = 2\pi r$ radius $r$ and the $\Phi = B\pi r^{2}$ value. At the Meissner effect observed at any radius $r$ of superconductor and any $B < B_{c1}$, this change is macroscopic $2\pi \hbar (- \Phi /\Phi_{0}) = 2\pi \hbar (-B \pi r^{2}/\Phi_{0}) \approx  -\hbar \ 10^{15}$ at the first critical field $B_{c1} \approx  0.1 \ T$ and the superconductor radius $r = 1 \ m$. This obscurity is macroscopic in truth because of the angular momentum change of all $N_{s} = n_{s}\pi r^{2}h > 10^{29}$ pairs in a cylindrical superconductor. Jorge Hirsch wonders fairly that {\it "the question of what is the 'force' propelling the mobile charge carriers and the ions in the superconductor to move in direction opposite to the electromagnetic force in the Meissner effect was essentially never raised nor answered"} \cite{Hirsch2010}. He proposes an explanation of the Meissner effect puzzle \cite{Hirsch2010}. But some consequences of this explanation, for example the electric field inside the superconductor, the relation (23) in \cite{Hirsch2010}, seem unacceptable.

According to the point of view by Hirsch \cite{Hirsch2003} the force-free momentum transfer indicates a fundamental problem with the conventional theory of superconductivity \cite{BCS1957}. I think that it indicates a fundamental problem rather with QM as a whole than with a theory of superconductivity \cite{PLA2012}. The persistent current $I_{p} \neq 0$ is observed at $R > 0$ not only in superconductor \cite{PCScien07,Letter07,toKulik2010} but also in normal metal rings \cite{PCScien09}. In order to dodge the obvious puzzle the authors \cite{PCScien09} and the author \cite{Birge2009} claim that the electric current can flow in realistic normal metal rings containing atomic defects, grain boundaries, and other kinds of static disorder without dissipating energy. They do not try even to explain how a dissipationless current of electrons can be possible at electron mean free path shorter than the circle length of their rings \cite{PCScien09}. The authors \cite{PCScien09} find a pretext for the dropping of the obvious puzzle using {\it a familiar analog in atomic physics: a current circulating around the atom} although the exponential decrease of the persistent current amplitude with temperature increase, Fig.3 in \cite{PCScien09}, testifies against this analog and to fundamental differences between application of some quantum principles on atomic and mesoscopic levels \cite{FFP8}. Igor Kulik, who has described the possibility of  $I_{p} \neq 0$ at $R > 0$ as far back as 1970 both in superconductor \cite{Kulik1970s} and normal metal rings \cite{Kulik1970n} made forty years ago reasonable statement that the taking into account of a dissipation should not result in the disappearance of the persistent current. The observation \cite{Letter07,toKulik2010} of the persistent current flowing against dc electric field confirms the Kulik's statement and makes meaningless \cite{toKulik2010} the preposterous claim by the authors \cite{PCScien09,Birge2009}. 

The authors \cite{PCScien09,Birge2009} turn a blind eye also to the other puzzle. It is known that {\it "time-reversal symmetry should forbid a current choosing one direction over the other around the ring"} \cite{Birge2009}. According to \cite{PCScien09,Birge2009} {\it "A magnetic flux $\Phi$ threading the ring will break time-reversal symmetry, allowing the PC to flow in a particular direction around the ring"} \cite{PCScien09}. But the $I_{p}$ \cite{PCScien07,PCScien09} and $E = -\bigtriangledown V_{p}$ \cite{Letter07,toKulik2010} direction changes not only with the $\Phi$ direction but also with its value at $\Phi = n\Phi_{0}$ and $\Phi = (n+0.5)\Phi_{0}$. Each physicist must understand that the observation of the direction change with the value change is a puzzle which may have a fundamental importance \cite{FQMT2004,QTRF2007,AIP2011PC}. Such puzzle was not observed on the atomic level because of the inaccessibly high magnetic fields $\Phi_{0}/ \pi r_{B}^{2} \approx 5 \ 10^{9} \ G$ needed for this. 

\subsection{Experimental results which can not be describe with help of the quantum formalism}
According to the criterion of demarcation between what is and is not genuinely scientific by Karl Popper: a theory should be considered scientific if and only if it is falsifiable. Therefore any experimental results contradicting to the quantum formalism should be at the centre of attention. I would like to draw reader's attention to two of such results. According to the quantum formalism (17) two permitted states $n$ and $n+1$ of superconducting loop should be observed at $\Phi = (n+0.5)\Phi_{0}$ at the single-shot measurement. Some experimental results \cite{1Shot02,1Shot04} corroborate this prediction but other one \cite{1Shot04} contradict to it. The $\chi $-shaped crossing observed in \cite{1Shot04} and misinterpreted \cite{QI2008} by the authors as the single-shot readout of macroscopic quantum superposition of  'flux qubit' \cite{Clarke2008,Mooij2003} states challenges the quantum formalism forbidding the state with $v = 0$ (17) at $\Phi = (n+0.5)\Phi_{0}$. Other challenge has been revealed at measurements of magnetic dependencies of the critical current of asymmetric superconducting rings \cite{PCJETP07,JETP07J,PRL06Rej,NANO2011}. According to the quantum formalism (17) the critical current anisotropy should appear in the asymmetric ring because of a change of the functions describing its magnetic dependencies, see Fig.19 \cite{PCJETP07} and  Fig.3 \cite{PRL06Rej}. But the measurements have revealed that the asymmetry appears because of changes in the arguments of the functions rather than the functions themselves \cite{PCJETP07,JETP07J,PRL06Rej}.

\section{Conclusion}
The history of QM demonstrates clearly that no successfulness of a theory can guarantee its accuracy. Most physicists believed in QM and disregarded the pointed criticism of Einstein, Schrodinger, Bell and others who tried to explain that the proposal of Heisenberg to describe observables instead of beables and the Born's interpretation of the wave function are inadequate. They strode unimpeded through the fundamental obscurity and non-universal validity quantum principles. Schrodinger pointed out, for example, the non-universal validity of the complementarity and uncertainty principles. According to Bohr {\it "the study of the complementary phenomena demands mutually exclusive experimental arrangements"} \cite{Bohr1949} But the method of momentum $p = mv$ measurement, learned in the primary school, disproves this statement. According to this method the momentum $p = m(z_{2}-z_{1})/(t_{2}-t_{1})$ is measured with help of measurement of the time $t_{1}$ and  $t_{2}$ when the particle passes points $z_{1}$ and $z_{2}$. Thus, the experimental arrangements for the measurement of position and momentum are not merely mutually non-exclusive but are the same. The velocity value $v_{z} = z/t$ can be measured with the uncertainty $\Delta v_{z} \approx v_{z}(\Delta z/z + \Delta t/t)$ at $z = z_{2} - z_{1} \gg  \Delta z$, $t = t_{2} - t_{1} \gg  \Delta t$. Consequently, we can make the product of the velocity $\Delta v_{z}$ and position $\Delta z$ uncertainties $\Delta z  \Delta v_{z} \approx \Delta z  v_{z}( \Delta z/z + \Delta t/t)  $ how any small, contrary to the Heisenberg uncertainty relation $\Delta z \Delta v_{z} > \hbar /2m$, increasing the distance $z = z_{2} - z_{1}$ and the time $t = z/v_{z}$. 

The mass delusion concerning quantum mechanic is a consequence of the optimistic point of view that the history of science is the history of progress. But it is also the history of wrong belief. In the beginning of his talk "Speakable and unspeakable in quantum mechanics" p. 169 in \cite{Bell2004} Bell cites a quotation from A. Koestler's book 'The Sleepwalkers' about of the Copernican revolution:  {\it "$\cdot \cdot \cdot $ the history of cosmic theories may without exaggeration be called a history of collective obsessions and controlled schizophrenias; and the manner in which some of the most important individual discoveries were arrived at reminds one of a sleepwalker's performance $\cdot \cdot \cdot $."}. According to Bell and Koestler Copernicus, Kepler, and Galilei {\it "were not really aware of what they were doing $\cdot \cdot \cdot $ sleepwalkers"} p. 169 in \cite{Bell2004}. This sleepwalking in the history of QM is more evident. QM is the most successful theory but it is also most obscure theory. The cause of this obscurity is obvious and some experts denoted it, for example Jaynes \cite{Jaynes1980}: {\it "From this, it is pretty clear why present quantum theory not only does not use - it does not even dare to mention - the notion of a 'real physical situation'. Defenders of the theory say that this notion is philosophically naive, a throwback to outmoded ways of thinking, and that recognition of this constitutes deep new wisdom about the nature of human knowledge. I say that it con-stitutes a violent irrationality, that somewhere in this theory the distinction between reality and our knowledge of reality has become lost, and the result has more the character of medieval necromancy than science"} see p. 231 in the book \cite{QuCh2006}.

But most physicists continue to believe in QM. No science can be possible without a faith. Scientists should, at least, believe in the human capability to perceive the outer world. But the belief should not be implicit. We should understand that the EPR correlation and violation of the Bell's inequalities rather cast doubt on the our capability to perceive the outer world than give new opportunities, for example quantum computation. Such sensible view of things hardly may be possible without the comprehension epistemological basic of science. Einstein emphasised this: {\it "The reciprocal relationship of epistemology and science is of noteworthy kind. They are dependent upon each other. Epistemology without contact with science becomes an empty scheme. Science without epistemology is - insofar as it is thinkable at all - primitive and muddled. However, no sooner has the epistemologist, who is seeking a clear system, fought his way through to such a system, than he is inclined to interpret the thought-content of science in the sense of his system and to reject whatever does not fit into his system. The scientist, however, cannot afford to carry his striving for epistemological systematic that far. He accepts gratefully the epistemological conceptual analysis; but the external conditions, which are set for him by the facts of experience, do not permit him to let himself be too much restricted in the construction of his conceptual world by the adherence to an epistemological system. He therefore must appear to the systematic epistemologist as a type of unscrupulous opportunist: he appears as realist insofar as he seeks to describe a world independent of the acts of perception; as idealist insofar as he looks upon the concepts and theories as the free inventions of the human spirit (not logically derivable from what is empirically given); as positivist insofar as he considers his concepts and theories justified only to the extent to which they furnish a logical representation of relations among sensory experiences. He may even appear as Platonist or Pythagorean insofar as he considers the viewpoint of logical simplicity as an indispensable and effective tool of his research"} \cite{Einstein1949}. 

The logical simplicity can not be attributable to QM. Moreover this theory is remarkable for logical inconsistency and non-universality. Even its fundamental obscurities are no-universal. The fundamental obscurity unmasked by Einstein, Schrodinger, Bell and other opponents of the positivism may be connected with the repudiation of realism. But even realistic description of many quantum phenomena with help of the Schrodinger's interpretation of the wave function has fundamental obscurities. We must conclude that a consistent and universal theory of quantum phenomena is absent now. "Could such theory be created in principle?" is the question requiring an answer first of all.


\begin{thebibliography}{99}
\bibitem{Bell2004} Bell J.S. Speakable and Unspeakable in Quantum Mechanics. Collected Papers on Quantum Philosophy. Cambridge University Press, Cambridge, 2004.

\bibitem{Heisenberg1925} Heisenberg W. Uber quantentheoretische Umdeutung kinematischer und mechanischer Beziehungen. Zeitschrift fur Physik 33, 879-893 (1925).

\bibitem{Bell1976} Bell J.S. The theory of local beables. Epistemological Letters, March 1976; p. 52-62 in [1]

\bibitem{Schrodinger1926} Schrodinger E. An Undulatory Theory of the Mechanics of Atoms and Molecules. Phys. Rev. 28, 1049-1070 (1926)

\bibitem{Schrodinger1952} Schrodinger E. Are there Quantum Jumps? Brit. J. Philos. Sci. 3, 233-242 (1952).

\bibitem{Bell1987} Bell J.S. Are there quantum jumps? In Schrodinger. Centenary of a polymath. Cambridge University Press (1987); p. 201-212 in [1].

\bibitem{WakeUp} Nikulov A.V. It is needed to shout 'wake up'. arXiv: 1008.5389 (2010)

\bibitem{Comment2010} Nikulov A.V. Comment on 'Coherent Detection of Electron Dephasing', arXiv: 1006.5662 (2010)

\bibitem{Comment2009} Nikulov A.V. Comment on 'Probing Noise in Flux Qubits via Macroscopic Resonant Tunneling'. arXiv: 0903.3575 (2009)

\bibitem{Aristov2011} V.V. Aristov and A.V. Nikulov, The fundamental obscurity in quantum mechanics. Why it is needed to shout "wake up", Proceedings of 19th International Symposium "NANOSTRUCTURES: Physics and Technology", Russia, St. Petersburg, Ioffe Physical-Technical Institute p. 145-146 (2011); arXiv: 1108.2628

\bibitem{Mooij2010} Mooij J.E. Quantum mechanics: No moon there. Nature Physics 6, 401-402, (2010)

\bibitem{COST} The COST Action MP1006 "Fundamental Problems in Quantum Physics". http://www.equantum.eu/

\bibitem{FeynmanL} Feynman R.P., Leighton R.B., Sands M. The Feynman Lectures on Physics, Addison-Wesley Publishing Company, Reading, Massachusetts, 1963.

\bibitem{Everett1957} Everett H. 'Relative State' Formulation of Quantum Mechanics. Rev. Mod. Phys. 29, 454-462 (1957)

\bibitem{Einstein1928L} A. Einstein, letter to E. Schrodinger, 31 May 1928, reprinted in Letters on Wave Mechanics, ed. M. Klein (New York: Philosophical Library, 1967).

\bibitem{QuCh2006} Greenstein G. and Zajonc A. The Quantum Challenge. Modern Research on the Foundations of Quantum Mechanics, 2nd edn. Jones and Bartlett, Sudbury 2006

\bibitem{BiomolInt03} Hackermuller L., Uttenthaler S., Hornberger K., Reiger E., Brezger B., Zeilinger A., and Arndt M. Wave Nature of Biomolecules and Fluorofullerenes, Phys. Rev. Lett. 91, 090408 (2003).

\bibitem{BiomolInt07}Gerlich  S., et al. A Kapitza-Dirac-Talbot-Lau interferometerfor highly polarizable molecules. Nature Physics 3,  711 - 715 (2007).

\bibitem{LL}  Landau L. D. and Lifshitz E. M. Quantum Mechanics: Non-Relativistic Theory. Volume 3, Third Edition, Elsevier Science, Oxford, 1977.

\bibitem{Zeilinger02} Nairz O., Arndt M., Zeilinger A. Experimental verification of the Heisenberg uncertainty principle for fullerene molecules. Phys. Rev. A 65, 032109 (2002).

\bibitem{SchrodingerHum} Schrodinger E. Science and Humanism. Physics in Our Time. Cambridge: University Press, 1952.   

\bibitem{Heisenberg1927} Heisenberg W. The Physical Content of Quantum Kinematics and Mechanics. Zeilschrift fur Physik, 43, 172-198 (1927).

\bibitem{Bohr1928} Bohr N. The Quantum Postulate and the Recent Development of Atomic Theory. Nature, 121, 580-90 (1928).  

\bibitem{EPR1935} Einstein A., Podolsky B. and Rosen N. Can quantum-mechanical description of physical reality be considered complete? Phys. Rev., 47, 777-780 (1935)

\bibitem{Bell1989} Bell J S, Against 'measurement'. in the proceedings of 62Years of Uncertainty (PlenumPublishing, New York 1989); p. 213 in  the book [1].

\bibitem{Kampen1988} van Kampen N.G. Ten theorems about quantum mechanical measurements. Physica A 153, 97-113 (1988)

\bibitem{Bell1981} Bell J.S. Bertlmann's socks and the nature of reality. Journal de Physique, 42, 41-61 (1981); p. 139 in the book [1].

\bibitem{Bohm1951} Bohm D.  Quantum Theory. New York: Prentice-Hall, 1951.

\bibitem{Dirac1930} Dirac A.M. Quantum Mechanics. Oxford University Press, 1930.

\bibitem{Neumann1932} von Neumann J. Mathematische Grundlagen der Quantenmechanik. Berlin: Springer, 1932; Mathematical Foundations of Quantum Mechanics. Princeton University Press, 1955.

\bibitem{Heisenberg1959} Heisenberg W. Physics and Philosophy. George Allen and Unwin Edition, 1959.

\bibitem{entanglement2001} Brukner C., Zukowski M., Zeilinger A., The essence of entanglement. arXiv: quant-ph/0106119

\bibitem{Schrod35D}  Schrodinger E. The present situation in quantum mechanics. Naturwissenschaften 23, 844-849 (1935); English translation published in Quantum Theory and Measurement edited by J.A. Wheeler and W.H. Zurek, Princeton University Press, Princeton, p. 152-167 (1983). 

\bibitem{Schrod35E} Schrodinger E. Discussion of probability relations between separated systems. Proc. Cambridge Phil. Soc. 31, 555-563 (1935).

\bibitem{Bell1990} Bell J.S. La nouvelle cuisine. in Between Science and Technology, (eds. A. Sarlemijn and P. Kroes) (Elsevier, North-Holland, 1990) p. 97-109; p. 232 in the book [1].

\bibitem{Bell1964} Bell J. S. On the Einstein-Podolsky-Rosen paradox. Physics 1, 195-200 (1964); p. 14 in the book [1].

\bibitem{EinsteiAutob} A. Einstein, Autobiography. in Albert Einstein, Philosopher Scientist, Edited by P. A. Schilp, p. 85, Library of Living Philosophers, Evanston, Illinois (1949).

\bibitem{Bell1982} Bell J.S. On the impossible pilot wave.  Found. Phys. 12, 989-999 (1982); p. 159 in the book [1].

\bibitem{StGe1922} Gerlach W. and Stern O. Das magnetische Moment des Silberatoms. Zs. Phys. 9, 353–355 (1922).

\bibitem{Bohr1949} Bohr N. Discussion with Einstein on epistemological problems in atomic physics. in Albert Einstein: Philosopher-Scientist, P. A. Schilpp, ed., pp. 200-41, The Library of Living Philosophers, Evanston (1949). 

\bibitem{Einstein1922} Einstein A. and Ehrenfest P. Quantentheoretische Bemerkungen zum Experiment von Stern und Gerlach. Zs. Phys. 11, 31-34 (1922).

\bibitem{Bell1966} Bell J.S. On the problem of hidden variables in quantum mechanics. Rev. Mod. Phys. 38, 447-452 (1966); p.1 in the book [1].

\bibitem{Mermin1993} Mermin N.D. Hidden variables and the two theorems of John Bell. Rev. Mod. Phys. 65, 803-815 (1993)

\bibitem{Adler1992} Adler Mortimer J. Natural Theology, Chance, and God. In The Great Ideas Today. Encyclopedia Britannica, Chicago, p. 300, 1992.  

\bibitem{Aspect1981} Aspect A., Grangier P., and Roger G. Experimental Tests of Realistic Local Theories via Bell's Theorem. Phys. Rev. Lett. 47, 460-463 (1981)

\bibitem{Aspect1982} Aspect A., Dalibard J., and Roger, G. Experimental Test of Bell's Inequalities Using Time - Varying Analyzers. Phys. Rev. Lett. 49, 1804-1807 (1982). 

\bibitem{Aspect82} Aspect A., Grangier P., and Roger G. Experimental Realization of Einstein-Podolsky-Rosen-Bohm Gedankenexperiment: A New Violation of Bell's Inequalities. Phys. Rev. Lett. 49, 91 - 94 (1982). 

\bibitem{Nielsen2000} Nielsen M.A. and Chuang I.L. Quantum Computation and Quantum Information. Cambridge University Press, 2000.

\bibitem{Mermin1985} Mermin N.D. Is the moon there when nobody looks? Reality and the quantum theory. Physics Today, 38, 38-47 (1985). 

\bibitem{Einstein1949} Einstein A. Remarks concerning the essays brought together in this co-operative volume. in Albert Einstein philosopherscientist, ed. by P.A. Schillp, Evanston, Illinois, pp. 665-688, 1949. 

\bibitem{Bohr1935EPR} Bohr N. Can Quantum-Mechanical Description of Physical Reality be Considered Complete? Phys. Rev. 48, 696-702 (1935)

\bibitem{Feynman1982} Feynman R. P. Int. J. Theor. Phys. 21 467 (1982)

\bibitem{FatherQC2007} Norton Q. The Father of Quantum Computing. http://www.wired.com/science/discoveries/news/2007/02/72734

\bibitem{Shor1994} Shor P. W. Algorithms for quantum computation: Discrete logarithms and factoring. in Proceedings of the 35th Annual Symposium on Foundations of Computer Science, IEEE Computer Society Press, Los Alamitos, CA, pp. 124-134; E- print: quantph/ 9508027

\bibitem{Deutsch1997FR} Deutsch D. The Fabric of Reality. The Penguin Press, 1997.

\bibitem{Heisenberg1969} Heisenberg W. Der Teil und das Ganze. Gesprache im Umkreis der Atomphysik. Munchen, 1969

\bibitem{electronInt} Tonomura A., Endo J., Matsuda T., Kawasaki T., and Exawa H. Demonstration of single-electron buildup of an interference pattern. Amer. J. Phys. 57, 117-120 (1989). 

\bibitem{AB1959} Aharonov Y. and Bohm D. Significance of Electromagnetic Potentials in the Quantum Theory. Phys. Rev. 115, 485 -491 (1959). 

\bibitem{ABRMP1985} Olariu S. and Popescu I. I. The quantum effects of electromagnetic fluxes. Rev. Mod. Phys. 57, 339 - 436 (1985)

\bibitem{PCScien09} Bleszynski-Jayich. A. C., et al.: Persistent Currents in Normal Metal Ring Science 326, 272-275 (2009).

 \bibitem{PCScien07} Koshnick N.C., Bluhm H., Huber M. E., Moler K.A. Fluctuation Superconductivity in Mesoscopic Aluminum Rings. Science 318, 1440-1443 (2007).

 \bibitem{PCJETP07} Gurtovoi V. L., Dubonos S. V., Nikulov A. V., Osipov N. N. and Tulin V. A. Dependence of the magnitude and direction of the persistent current on the magnetic flux in superconducting rings. J. of Experimental and Theoretical Physics 105, 1157-1173 (2007); arXiv: 0903.3539

\bibitem{ABNanoSt2009} Gurtovoi V. L., Nikulov A. V., and Tulin V. A. Aharonov-Bohm effects in nanostructures. in Proceedings of 17th International Symposium "NANOSTRUCTURES: Physics and Technology", Belarus, Minsk, Institute of Physics NAS, p. 87 (2009); arXiv: 0910.5172

\bibitem{FPP2008} Nikulov A. V. Bohm's quantum potential and quantum force in superconductor. AIP Conference Proceedings, Vol. 1101 "Foundations of Probability and Physics-5" pp. 134-143 (2009); arXiv: 0812.4118.

\bibitem{Strambini} Strambini E., et al. Coherent Detection of Electron Dephasing. Phys. Rev. Lett. 104, 170403 (2010).

\bibitem{Nature08}Tonomura A. and Nori, F. Disturbance without the force. Nature 452, 298-300 (2008).

\bibitem{Jammer74} Jammer M. The Philosophy of Quantum Mechanics, John Wiley (1974), p. 204, quoting A. Petersen, Bulletin of the Atomic Scientist 19, 12 (1963). 

\bibitem{Korotkov2010}  Palacios-Laloy A., et al. Experimental violation of a Bell's inequality in time with weak measurement. Nature Physics 6, 442-447, (2010)

\bibitem{Clarke2008} Clarke J. and Wilhelm F.K. Superconducting quantum bits. Nature 453, 1031-1042 (2008).

\bibitem{Leggett1985} Leggett A.J. and Garg A. Quantum mechanics versus macroscopic realism: Is the flux there when nobody looks? Phys. Rev. Lett. 54, 857-860 (1985)

\bibitem{Mooij2003} Chiorescu I., Nakamura Y., Harmans C.J.P.M., and Mooij J.E., Coherent Quantum Dynamics of a Superconducting Flux Qubit. Science 299, 1869-1871 (2003)

\bibitem{Nikulov2010} Nikulov A.V. Flux-qubit and the law of angular momentum conservation. Quantum Computers and Computing. 10, 42-61 (2010); arXiv:1005.3776.

\bibitem{Leggett2002} Leggett A.J. Superconducting Qubits - a Major Roadblock Dissolved? Science 296, 861-862 (2002)

\bibitem{JLTP1998} Nikulov A.V., Zhilyaev I.N. The Little-Parks Effect in an  Inhomogeneous Superconducting Ring. J. Low Temp.Phys. 112, 227-236 (1998).

\bibitem{PCScien07} Koshnick N.C., Bluhm H., Huber M. E., and Moler K.A. Fluctuation Superconductivity in Mesoscopic Aluminum Rings. Science 318, 1440-1443 (2007).

\bibitem{Tinkham}  Tinkham M. Introduction to Superconductivity. McGraw-Hill Book Company, New-York, 1975.

\bibitem{PhysLetA2012} Burlakov A.A., Gurtovoi V. L., Ilin A. I., Nikulov A. V., and Tulin V. A. A possibility of persistent voltage observation in a system of asymmetric superconducting rings. Phys. Lett. A 376, 2325-2329 (2012); arXiv: 1207.0791 

\bibitem{toKulik2010} Gurtovoi V.L., Ilin A.I., Nikulov A.V., and Tulin V.A. Weak dissipation does not result in disappearance of persistent current. Fizika Nizkikh Temperatur, 36, 1209-1218 (2010); arXiv: 1004.3914

\bibitem{Letter07} Burlakov A. A., Gurtovoi V.L., Dubonos S.V., Nikulov A.V., and Tulin V.A. Little-Parks Effect in a System of Asymmetric Superconducting Rings. JETP Letters 86, 517-521 (2007); arXiv: 0805.1223.

\bibitem{PerMob2001} Dubonos S.V., Kuznetsov V.I., and Nikulov A.V. Segment of an Inhomogeneous Mesoscopic Loop as a DC Power Source. in Proceedings of 10th International Symposium "NANOSTRUCTURES: Physics and Technology" St Petersburg: Ioffe Institute, p. 350-352 (2002); arXiv: physics/0105059

\bibitem{Letter2003} Dubonos S. V., Kuznetsov V. I., Zhilyaev I. N., Nikulov A. V., and Firsov A. A.: Observation of the External-ac-Current-Induced dc Voltage Proportional to the Steady Current in Superconducting Loops. Pis'ma Zh. Eksp. Teor. Fiz. 77, 439-444  (2003) (JETP Letters, 77, 371-375 (2003)); arXiv: cond-mat/0303538.

\bibitem{LP1962} Little W. A. and Parks R. D. Observation of Quantum Periodicity in the Transition Temperature of a Superconducting Cylinder. Phys. Rev. Lett. 9,  9-12 (1962)

\bibitem{PRB2001} Nikulov A.V. Quantum Force in Superconductor. Phys.Rev. B 64, 012505 (2001).

\bibitem{Meissner1933} Meissner W. and Ochsenfeld R. Ein neuer effekt bei eintritt der supraleitfahigkeit. Naturwissenschaften 21, 787-788 (1933).

\bibitem{Hirsch2010} Hirsch, J. E.: Electromotive forces and the Meissner effect puzzle. J. of Superconductivity and Novel Magnetism {\bf 23}, 309-317 (2010); arXiv: 0908.409

\bibitem{Hirsch2003} Hirsch J. E.  The Lorentz force and superconductivity. Phys.Lett. A 315, 474-477 (2003); arXiv: cond-mat/0305542

\bibitem{BCS1957} Bardeen J., Cooper L. N., and Schrieffer J. R. Theory of Superconductivity. Phys. Rev. 108, 1175-1205 (1957).

\bibitem{PLA2012} Nikulov A.V. The Meissner effect puzzle and the quantum force in superconductor. Phys. Lett. A 376, 3392-3397 (2012); arXiv:1211.0002

\bibitem{Birge2009} Birge N. O. Sensing a Small But Persistent Current. Science 326, 244 - 245 (2009).

\bibitem{FFP8} Nikulov A.V. Fundamental Differences Between Application of Basic Principles of Quantum Mechanics on Atomic and Higher Levels. AIP Conference Proceedings, Vol. 905, FRONTIERS OF FUNDAMENTAL PHYSICS: Eighth International Symposium, pp. 117-119 (2007); arXiv: 0707.0827.

\bibitem{Kulik1970s} Kulik I.O. Magnetic flux quantization in the normal state. Zh. Exsp. Teor. Fiz. 58, 2171-2175 (1970) (Sov. Phys. JETP 31, 1172-1176 (1970)) 

\bibitem{Kulik1970n} Kulik I.O. Flux quantization in normal metal. Pis'ma Zh. Eksp. Teor. Fiz. 11, 407 (1970) (JETP Lett. 11, 275 (1970)).

\bibitem{FQMT2004} Nikulov A. V. Quantum limits to the second law and breach of symmetry. arXiv: cond-mat/0505508. 

\bibitem{QTRF2007} Nikulov A.V. About Essence of the Wave Function on Atomic Level and in Superconductors. AIP Conference Proceedings, Vol.  962, QUANTUM THEORY: Reconsideration of Foundations - 4, pp. 297-301 (2007); arXiv: 0803.1840.

\bibitem{AIP2011PC} Nikulov A.V. Observations of Persistent Current at Non-Zero Resistance: Challenge to the Second Law of Thermodynamics. AIP Conference Proceedings, 
 Vol. 1411, Second Law of Thermodynamics: Status and Challenges, pp. 122-144 (2011)  

\bibitem{1Shot02} Tanaka H., Sekine Y., Saito S., and Takayanagi H. DC-SQUID readout for qubit. Physica C 368, 300-304 (2002). 

\bibitem{1Shot04} Tanaka H., Saito S., Nakano H., Semba K., Ueda M., and Takayanagi H.  Single-Shot Readout of Macroscopic Quantum Superposition State in a Superconducting Flux Qubit. arXiv: cond-mat/0407299.

\bibitem{QI2008} Aristov V. V. and  Nikulov A. V. Quantum computation and hidden variables. Proceedings SPIE Vol. 7023, Quantum Informatics 2007 (2008).

\bibitem{JETP07J} Gurtovoi V.L., Dubonos S.V., Karpi S.V., Nikulov A.V., and Tulin V.A. Contradiction between the Results of Observations of Resistance and Critical Current Quantum Oscillations in Asymmetric Superconducting Rings. J. of Experimental and Theoretical Physics, 105, 257-262 (2007); arXiv: 1011.6149.

\bibitem{PRL06Rej} Burlakov A. A., Gurtovoi V.L., Dubonos S.V., Nikulov A.V., and Tulin V.A. Could two degenerate energy states be observed for a superconducting ring at $\Phi_{0}$/2? arXiv: cond-mat/0603005

\bibitem{NANO2011} Burlakov A.V., Gurtovoi V.L., Ilin A.I., Nikulov A.V., and Tulin V.A. Experimental investigations of the change with magnetic flux of quantum number in superconducting ring. Proceedings of 19th International Symposium "NANOSTRUCTURES: Physics and Technology", Russia, St. Petersburg, Ioffe Physical-Technical Institute p. 128-129 (2011); arXiv: 1103.3115

\bibitem{Jaynes1980} Jaynes E.T. Quantum beats. in Foundations of Radiation Theory and Quantum Electrodynamics, A.O. Barut (ed.), New York: Plenum Press, 1980.

\end{thebibliography}
\end{document}